\documentclass[a4paper,11pt]{article}
\pdfoutput=1 

\usepackage{jheppub} 
\usepackage{slashed}
\usepackage[T1]{fontenc}
\usepackage{fix-cm}
\usepackage{lmodern}
\usepackage{amsmath,amssymb,mathtools,mathrsfs,empheq}
\usepackage{physics}
\usepackage{bbm}
\usepackage{bm}
\usepackage{hhline}
\usepackage{cancel}
\preprint{CQUeST-2026-0803}

\newcommand\fft[2]{\frac{#1}{#2}}
\newcommand\nn{\nonumber}

\newcommand\mO{\mathcal{O}}
\newcommand\tir{\tilde{r}}

\newcommand\mD{\mathcal{D}}
\newcommand\mN{\mathcal{N}}

\newcommand\tmu{\tilde{\mu}}

\newcommand\bh{\bar{h}}

\newcommand\mR{\mathcal{R}}

\newcommand\mX{\mathcal{X}}

\newcommand\SU{\text{SU}}

\title{\boldmath Anatomy of AdS black hole on-shell actions}

\author{Junho Hong}
\author{and Sourav Roychowdhury}

\affiliation{Department of Physics \& Center for Quantum Spacetime, Sogang University \\ 35 Baekbeom-ro, Mapo-gu, Seoul 04107, Republic of Korea }

\emailAdd{junhohong@sogang.ac.kr}
\emailAdd{srcphys@sogang.ac.kr}

\abstract{We revisit the regularized on-shell actions of rotating electrically charged AdS black holes of gauged supergravity via holographic renormalization and background subtraction, with a particular emphasis on the frame in which the radial cutoff is placed. For holographic renormalization, the asymptotically global AdS frame, recently employed in seven dimensions, works equally well in five and six dimensions: the renormalized on-shell action of the AdS$_d$ black holes with $S^1\times S^{d-2}$ conformal boundaries satisfies the quantum statistical relation, up to a parameter-independent Casimir term in odd dimensions, and thereby places the holographic comparisons that so far relied on this relation on a first-principles footing. Background subtraction, in contrast, fails to reproduce this result in the same frame; the mismatch, which we identify exactly in terms of the Casimir term and the counterterms, demands a criterion for the cutoff surface yet to be established. For precision holography, our results settle on a convenient choice of frame for holographic renormalization of these black holes and delineate the ambiguities that remain in background subtraction.}
\begin{document} 
\maketitle
\flushbottom

\section{Introduction}\label{sec:intro}

Black holes are central objects of theoretical physics, and the Euclidean gravitational path integral has been a particularly fruitful tool in deciphering their physics. Most prominently, the Bekenstein--Hawking entropy and the laws of black hole thermodynamics \cite{Bekenstein:1973ur,Bardeen:1973gs,Hawking:1975vcx} are encoded in a thermodynamic potential obtained from the on-shell action of the Euclidean black hole saddle, which governs the semi-classical limit of the gravitational path integral \cite{Gibbons:1976ue}. In the same spirit, competing saddles of the path integral are interpreted as competing thermodynamic phases, an interpretation on which the Hawking--Page transition between thermal AdS and the AdS black hole \cite{Hawking:1982dh}, its holographic reading as the deconfinement transition of the dual gauge theory \cite{Witten:1998zw,Sundborg:1999ue,Aharony:2003sx}, and the phase diagrams of charged and rotating AdS black holes \cite{Chamblin:1999tk,Chamblin:1999hg,Cvetic:1999ne,Hawking:1998kw,Caldarelli:1999xj} all rest. 

Within this framework, AdS black holes of gauged supergravity are of particular interest because of their role in holography. Through the AdS/CFT correspondence, their Euclidean on-shell action, the leading semi-classical approximation to the gravitational path integral, is identified with the `free energy' of the dual field theory on the conformal boundary. This identification is tested most sharply in the supersymmetric limit, where the dual free energy is encoded in a supersymmetric index \cite{Romelsberger:2005eg,Kinney:2005ej,Bhattacharya:2008zy,Benini:2015noa} computable exactly by localization \cite{Pestun:2007rz,Pestun:2016zxk}: the agreement of the on-shell action with minus the logarithm of the index at large $N$ has become a benchmark of precision holography, established for electrically charged rotating black holes in AdS$_4$ \cite{Bobev:2019zmz,Choi:2019zpz,Nian:2019pxj,Choi:2019dfu,Bobev:2022wem}, AdS$_5$ \cite{Cabo-Bizet:2018ehj,Choi:2018hmj,Benini:2018ywd}, AdS$_6$ \cite{Choi:2019miv,Crichigno:2020ouj}, and AdS$_7$ \cite{Choi:2018hmj,Nahmgoong:2019hko,Kantor:2019lfo,Bobev:2023bxl}, as well as for magnetically charged black holes in AdS$_4$ \cite{Benini:2015eyy,Benini:2016rke,Bobev:2020pjk} and AdS$_6$ \cite{Hosseini:2018uzp,Crichigno:2018adf,Suh:2018tul,Hosseini:2018usu,Suh:2018qyv,Fluder:2019szh}. The agreement carries a further lesson: the logarithm of the large $N$ index, regarded as a function of the chemical potentials, serves as an entropy function in that the extremum of its Legendre transform, subject to the supersymmetric constraint on the chemical potentials, reproduces the Bekenstein--Hawking entropy \cite{Benini:2015eyy,Benini:2016rke,Hosseini:2017mds,Hosseini:2018dob,Choi:2018fdc,Cabo-Bizet:2018ehj,Benini:2018ywd,Cassani:2019mms,Hosseini:2019iad}; see \cite{Zaffaroni:2019dhb,Cassani:2025sim} for reviews. This line of work thus opens an avenue towards a microscopic understanding of black holes, to be completed beyond the semi-classical limit.

Behind these successes lies a step that is easily taken for granted: any such comparison presupposes a finite on-shell action, whereas the on-shell action of an AdS black hole diverges with the infinite volume of AdS. It therefore has to be regularized, and there are two standard procedures, which will be reviewed in the main text: holographic renormalization and background subtraction. In holographic renormalization \cite{Henningson:1998gx,Balasubramanian:1999re,Emparan:1999pm,deHaro:2000vlm,Bianchi:2001kw,Skenderis:2002wp} the divergences are cancelled by covariant counterterms built from the fields induced on a cutoff surface, which is then sent to infinity. In background subtraction \cite{Gibbons:1976ue,Hawking:1982dh,Hawking:1995fd,Witten:1998zw} the on-shell action of a reference background, thermal AdS for the black holes at hand, is subtracted in place of the counterterms, with the fields of the reference suitably matched to those of the black hole on the cutoff surface.

Either procedure is available in principle, yet for the general non-extremal AdS black holes of gauged supergravities in five and more dimensions the on-shell action has mostly been inferred from the thermodynamic quantities through the quantum statistical relation (QSR) \cite{Chow:2007ts,Choi:2018fdc,Cassani:2019mms,Kantor:2019lfo,Bobev:2023bxl} rather than evaluated directly; direct evaluations beyond the Kerr--AdS \cite{Hawking:1998kw,Awad:2000aj,Gibbons:2004ai,Papadimitriou:2005ii} and static charged \cite{Chamblin:1999tk,Cvetic:1999ne,Batrachenko:2004fd} black holes had been confined to five dimensions, to the black hole of minimal gauged supergravity \cite{Kunduri:2005zg,Chen:2005zj,Bobev:2022ocx,Awad:2022exw} and the equal-rotation black hole of the STU model \cite{Cassani:2019mms,Ntokos:2021duk}, until the recent holographic renormalization of the AdS$_7$ black hole \cite{Bobev:2025xan}.   In this work we close this gap for the general non-extremal electrically charged rotating AdS black holes of gauged supergravity: we evaluate their on-shell actions by holographic renormalization in five and six dimensions and by background subtraction in five, six, and seven dimensions. Among the reasons why such a direct evaluation matters, we would like to highlight two.
\begin{itemize}
	\item Precision holography rests on the identification of the string/M-theory path integral with that of the dual CFT, and both are naturally formulated in Euclidean signature. For AdS black holes, the comparison must therefore be achievable on the Euclidean black saddle alone \cite{Bobev:2020pjk}; the QSR, in contrast, imports Lorentzian data, such as the entropy tied to the horizon of a black hole, into the on-shell action. The direct evaluation removes this detour and confirms that the Euclidean black saddle by itself suffices for precision holography.
	\item Equivariant localization \cite{BenettiGenolini:2023kxp,Martelli:2023oqk}, reviewed in \cite{Sparks:2026taw}, yields the regularized on-shell action of a supersymmetric solution from fixed-point data alone, as established for four- and six-dimensional gauged supergravity \cite{BenettiGenolini:2023kxp,BenettiGenolini:2024lbj,Couzens:2025ghx} and extended to five dimensions \cite{Colombo:2025ihp,BenettiGenolini:2025icr,Colombo:2025yqy,Park:2025fon,Cassani:2026teb,BenettiGenolini:2026cdw} and to higher-derivative corrections \cite{Hristov:2024cgj,BenettiGenolini:2026qdm,Gaar:2026nqq}. Hence one might conclude that the direct evaluation has become obsolete, but this is not the case. Localization neither dictates the choice between holographic renormalization and background subtraction, as discussed in \cite{BenettiGenolini:2025icr,Colombo:2025ihp}, nor resolves the subtleties within each, as we spell out in this paper. Moreover, it relies on supersymmetry, which the generic non-extremal black holes of interest do not have, whereas the holographic identification of the path integrals is not restricted to supersymmetric backgrounds. Our analysis of both regularizations for generic non-extremal black holes addresses precisely the ambiguity that localization leaves open, and does so without supersymmetry.
\end{itemize}

The paper is organized accordingly. In Section~\ref{sec:holo-renorm} we carry out holographic renormalization for the general rotating electrically charged AdS$_5$ black hole of the $\mathrm{U}(1)^3$ STU model and the AdS$_6$ black hole of $\mathcal N=4$ $\mathrm{SU}(2)$ gauged supergravity in the asymptotically global AdS frame, the frame in which the AdS$_7$ black hole was first renormalized in \cite{Bobev:2025xan}. The renormalized on-shell action satisfies the QSR exactly in six dimensions and, as in seven, up to a parameter-independent Casimir term in five, so that this frame works universally. In Section~\ref{sec:bs} we show that background subtraction at a constant radial coordinate of the same frame reproduces neither the renormalized on-shell action nor the on-shell action determined via the QSR in five, six, and seven dimensions; the mismatch, identified exactly in terms of the renormalized on-shell action of the reference and the finite parts of the counterterms, implies that background subtraction requires a criterion fixing the cutoff surface if it is to serve as an unambiguous regularization method for precision holography. Section~\ref{sec:discussion} discusses open questions, and the appendices collect our conventions and the integrals used in the main text.

\section{Holographic Renormalization}\label{sec:holo-renorm}

In this section we compute the renormalized Euclidean on-shell actions of the general rotating electrically charged, asymptotically AdS$_5$ and AdS$_6$ black holes of gauged supergravity by holographic renormalization: the AdS$_5$ black hole of $\mathcal N=2$ $\mathrm{U}(1)^3$ STU gauged supergravity in Subsection~\ref{sec:holo-renorm:AdS5} and the AdS$_6$ black hole of $\mathcal N=4$ $\mathrm{SU}(2)$ gauged supergravity in Subsection~\ref{sec:holo-renorm:AdS6}. Through this investigation, we find that holographic renormalization implemented in the asymptotically global AdS frame yields a renormalized on-shell action consistent with the quantum statistical relation (QSR) of black hole thermodynamics, up to a scheme-dependent Casimir term in odd dimensions. Our conventions for differential forms, the Hodge dual, and the Wick rotation are collected in Appendix~\ref{app:conventions}.

\subsection{AdS$_5$ black hole in $\mN=2$ STU gauged supergravity}\label{sec:holo-renorm:AdS5}


\noindent {\bf{Theory.}} The five-dimensional theory of interest is the $\mathrm{U}(1)^3$ truncation of the maximal $\mathrm{SO}(6)$ gauged supergravity, often called the STU model: $\mathcal N=2$ gauged supergravity coupled to two vector multiplets, whose three $\mathrm{U}(1)$ gauge fields generate the Cartan subgroup $\mathrm{U}(1)^3\subset\mathrm{SO}(6)$. It is a consistent truncation of type IIB supergravity on $S^5$ \cite{Cvetic:1999xp}, contained in the consistent $\mathrm{SO}(6)$ reduction of \cite{Cvetic:2000nc} and in the full truncation to maximal $\mathrm{SO}(6)$ gauged supergravity \cite{Baguet:2015sma}, so that every solution of the five-dimensional theory lifts to ten dimensions. The bosonic Lorentzian action reads (see \textit{e.g.}\ \cite{Cvetic:2004ny,Wu:2011gq,Cassani:2019mms})\footnote{The Hodge star operator follows the convention of Appendix~\ref{app:conventions}, and the orientation of the totally anti-symmetric symbol \eqref{eq:epsilon} that enters it is fixed, in the Boyer--Lindquist-type coordinates of \eqref{eq:metric5}, by $\epsilon_{tr\theta\phi_1\phi_2}=\sqrt{-g}$. The Hodge-dual terms of the action are insensitive to the orientation, whereas the Chern--Simons term is not; for the solution below, the orientation thus fixes the sign of the Chern--Simons term relative to the Hodge-dual terms, which here coincides with that of \cite[(2.1)]{Cassani:2019mms}.}
\begin{align}
	S_{\rm bulk}^L &=  \frac{1}{16 \pi G_N} \int \bigg[\Big(R-\mathcal V\Big) \star_L 1 - \frac{1}{2} \sum_{i=1}^2  \star_L  d\varphi_i \wedge d\varphi_i  \nn\\
	& \kern6.5em - \frac{1}{2} \sum_{I=1}^3 X_I^{-2} \star_L F_{(2)}^I \wedge F_{(2)}^I - \frac{1}{6} |\epsilon_{IJK}| F_{(2)}^I \wedge F_{(2)}^J \wedge A_{(1)}^K\bigg] \,, \label{eq:action5L}
\end{align}
where the superscript $L$ denotes Lorentzian signature, with mostly-plus metric.  The two scalars $\varphi_{1,2}$ are parametrized as
\begin{align}
	X_1 = e^{-\frac{\varphi_1}{\sqrt 6} - \frac{\varphi_2}{\sqrt 2} }\,,\qquad   X_2 = e^{-\frac{\varphi_1}{\sqrt 6} + \frac{\varphi_2}{\sqrt 2} }\,,\qquad  X_3 = e^{\frac{2\varphi_1}{\sqrt 6}} \,, \label{eq:param5}
\end{align}
the field strengths are
\begin{align}
	F_{(2)}^I = dA_{(1)}^I \,, \label{eq:F5}
\end{align}
and the scalar potential is written in terms of the superpotential $\mathcal W$ as \cite{Batrachenko:2004fd,Skenderis:1999mm,DeWolfe:1999cp,Cassani:2019mms}
\begin{align}
	\mathcal V = -4g^2 \sum_{I=1}^3 X_I^{-1} = 2\sum_{i=1}^2 \big(\partial_{\varphi_i}\mathcal W\big)^2 - \frac{4}{3}\,\mathcal W^2\,, \qquad \mathcal W = g \sum_{I=1}^3 X_I\,, \label{eq:potential5}
\end{align}
where $g$ is the gauge coupling parameter. The superpotential is stationary at the supersymmetric AdS$_5$ vacuum $X_I=1$ \cite{Skenderis:1999mm}, where $\mathcal W=3g$ and $\mathcal V=-12g^2$. 

Varying the action \eqref{eq:action5L} yields the Einstein equation
\begin{align}\label{eq:Einstein5}
	R_{\mu \nu} - \frac{1}{2} g_{\mu \nu} \big(R - \mathcal V\big) &= \sum_{i=1}^2  \frac{1}{2} \bigg[ \partial_\mu \varphi_i \partial_\nu \varphi_i - \frac{1}{2} g_{\mu \nu} g^{\rho \sigma} \partial_\rho \varphi_i \partial_\sigma \varphi_i \bigg] \nn\\
	&\quad + \sum_{I=1}^3  X_I^{-2} \frac{1}{2}  \bigg[ F^{I}_{\mu \rho} F^{I}_{\nu}{}^{\rho}  - \frac{1}{4} g_{\mu \nu} F^I_{\rho \sigma} F^{I\rho \sigma}\bigg] \,,
\end{align}
the scalar field equations
\begin{align}
	d(\star_L d \varphi_i) = \Bigg[\frac{1}{4} \sum_{I=1}^3 \partial_{\varphi_i}(X_I^{-2}) F_{\mu \nu}^I F^{I \mu \nu} + \partial_{\varphi_i} \mathcal{V}\Bigg] \star_L 1 \,, \label{eq:scalarEOM5}
\end{align}
and the Maxwell equations
\begin{align}
	d\big(X_I^{-2} \star_L F_{(2)}^I \big) + \frac{1}{2} |\epsilon_{IJK}| F_{(2)}^J \wedge F_{(2)}^K = 0 \, . \label{eq:Maxwell5}
\end{align}

\medskip

\noindent {\bf{Black hole solution.}} The general non-extremal, asymptotically AdS$_5$ black hole of this theory carries two independent rotation parameters and three independent $\mathrm{U}(1)$ electric charges \cite{Wu:2011gq}; it contains the two-rotation, equal-charge solution of minimal gauged supergravity \cite{Chong:2005hr} and the equal-rotation, three-charge solution \cite{Cvetic:2004ny} as special cases (see \cite{Bobev:2022ocx,Awad:2022exw,Cassani:2019mms} for their on-shell actions). Its local form in Boyer--Lindquist-type coordinates $(t,r,\theta,\phi_i)$ reads
\begin{align}\label{eq:metric5}
	ds_5^2 &= (H_1 H_2 H_3)^{\frac{1}{3}} \bigg[- \frac{(1+g^2r^2)\Delta_\theta}{\Xi_1 \Xi_2} dt^2+ \Sigma \Big(\frac{r^2}{\Delta_r} dr^2+\frac{1}{\Delta_\theta} d\theta^2\Big)+ \sum_{i=1}^2 \frac{r^2+a_i^2}{\Xi_i} \mu_i^2 d\phi_i^2 \nn\\
	& \kern7em + \sum_{I=1}^3\frac{2ms_I^2}{\Sigma H_I(s_I^2-s_J^2)(s_I^2-s_K^2)} K_I^2\bigg] \, ,  
\end{align}
where $(\mu_1,\mu_2)=(\sin\theta,\cos\theta)$ and $\{I,J,K\}$ is a permutation of $\{1,2,3\}$. In the metric \eqref{eq:metric5}, we have used
\begin{subequations}
\begin{align}
	s_I &=\sinh\delta_I\,,\qquad c_I = \cosh\delta_I\,,\qquad\Xi_i = 1-a_i^2g^2\,,\qquad \Xi_{Ii} = c_I^2 -s_I^2 \Xi_i\,,\\
	\Sigma&= r^2+\sum_{i=1}^2a_i^2(1-\mu_i^2)\,,\qquad \Delta_\theta = 1-g^2\sum_{i=1}^2a_i^2(1-\mu_i^2) \,,\\
	H_I &= 1 + \frac{2ms_I^2}{\Sigma}\,,	
\end{align}\label{eq:metric5def}%
\end{subequations}
the one-forms $K_I$ with constant coefficients $\kappa_I$ and $\kappa_{Ii}$,
\begin{align}
	K_I &= \frac{\Delta_\theta}{\Xi_1 \Xi_2}\, \kappa_I\, dt + \sum_{i=1}^2 \frac{\mu_i^2}{\Xi_i}\, \kappa_{Ii}\, d\phi_i\,, \label{eq:oneforms5}\\
	\kappa_I &= \frac{c_I^2}{c_1 c_2 c_3} P_I - \frac{g^2 a_1 a_2\, c_1 c_2 c_3}{c_I^2} N_I\,, \nn\\
	\kappa_{Ii} &= -\frac{a_i\,\Xi_{Ii}}{\sqrt{\Xi_{1i} \Xi_{2i} \Xi_{3i}}}P_I + \frac{a_1 a_2}{a_i}\frac{\sqrt{\Xi_{1i} \Xi_{2i} \Xi_{3i}}}{\Xi_{Ii}} N_I\,, \nn\\
	P_I &\equiv \frac{s_I c_1 c_2 c_3}{c_I} \frac{\sqrt{\Xi_{11} \Xi_{21} \Xi_{31} \Xi_{12} \Xi_{22} \Xi_{32}}}{\sqrt{\Xi_{I1} \Xi_{I2}}}\,, \qquad N_I \equiv \frac{c_I s_1 s_2 s_3}{s_I} \sqrt{\Xi_{I1} \Xi_{I2}}\,, \nn
\end{align}
and the radial function
\begin{align}
	\Delta_r &= \big(r^2+a_1^2\big) \big(r^2+a_2^2\big) \big(1+g^2r^2\big) - 2mr^2 + 2mg^2 \Big[s_A r^4 \nn\\
	&\quad - s_B \Big(\big(a_1^2+a_2^2- 2m\big) r^2+a_1^2 a_2^2 \big(2+g^2r^2\big)\Big) \nn\\
	&\quad + s_C\Big(\big((a_1+a_2)^2-2m\big)\big((a_1-a_2)^2-2m\big) -2g^2 a_1^2 a_2^2 \big(2r^2+2m+a_1^2+a_2^2\big)+g^4 a_1^4 a_2^4\Big) \nn\\
	&\quad +2mg^2 a_1^2 a_2^2 \big(s_B^2 -4 s_A s_C\big)\Big]\,,\label{eq:Delta5}
\end{align}
written compactly in terms of
\begin{align}
	s_A = \sum_{I=1}^3 s_I^2\,,\qquad s_B = \sum_{I<J} s_I^2 s_J^2\,,\qquad s_C = s_1^2 s_2^2 s_3^2\,.\label{eq:sABC5}
\end{align}
The black hole horizon is at $r=r_+$, the largest real root of the radial function \eqref{eq:Delta5}. It is a Killing horizon generated by the Killing vector $\xi^L=\partial_t+\Omega_i\partial_{\phi_i}$, where $\Omega_i$ are the angular velocities of the horizon presented in \eqref{eq:thermo5}. The scalars are\footnote{These are the inverses of the scalars $X_i$ in \cite{Wu:2011gq}.}
\begin{align}
	X_I = \frac{\big(H_1 H_2 H_3\big)^{\frac{1}{3}}} {H_I}\,,\label{eq:X5}
\end{align}
and the gauge fields are
\begin{align}
	A_{(1)}^I = \frac{2m}{\Sigma H_I} K_I + \alpha^I dt\,,\label{eq:A5}
\end{align}
where the $\alpha^I$ are constants: $\alpha^Idt$ is a flat connection, which drops out of the field strengths and of the equations of motion, and is fixed by regularity of the Euclidean solution below.

Note that in the above description of a black hole solution we have employed the angular coordinates for rotating black holes in general dimension \cite{Gibbons:2004uw,Gibbons:2004js}, consisting of the $n$ azimuthal angles $\phi_i$ of period $2\pi$ and the $n+\epsilon$ direction cosines $\mu_i$ obeying
\begin{align}
	\sum_{i=1}^{n+\epsilon} \mu_i^2 = 1\,, \qquad 0\le\mu_i\le1 \ \ (1\le i\le n)\,, \qquad -1\le\mu_{n+1}\le1 \ \ (\text{even } d)\,,\label{eq:mu}
\end{align}
where the dimension $d$ of the background is decomposed as $d=2n+\epsilon+1$ with $\epsilon=(d-1)\bmod2$, \textit{i.e.}\ $n=\lfloor(d-1)/2\rfloor$. These angular coordinates $(\mu_i,\phi_i)$ parametrize the unit sphere $S^{d-2}$, while $(r,\mu_i,\phi_i)$ become spheroidal coordinates. For the AdS$_5$ black hole \eqref{eq:metric5}, $(n,\epsilon)=(2,0)$ and we set $(\mu_1,\mu_2)=(\sin\theta,\cos\theta)$ with $0\le\theta\le\pi/2$; the same coordinates with $(n,\epsilon)=(2,1)$ and $(n,\epsilon)=(3,0)$ will be used to describe the AdS$_6$ and AdS$_7$ black holes of Subsection~\ref{sec:holo-renorm:AdS6} and Subsection~\ref{sec:bs:AdS7}.

\medskip

\noindent{\bf{Euclidean saddle.}} The Euclidean saddle of interest is obtained from the black hole solution above by the Wick rotation
\begin{align}
	t \rightarrow -i \tau\,. \label{eq:Wick}
\end{align}
Global regularity of the resulting Euclidean geometry fixes the thermal cycle and the flat connection of the gauge fields as follows.
\begin{itemize}
	\item The Euclidean time circle shrinks to zero size at the horizon $r=r_+$. The absence of a conical singularity there requires the identification \cite{Hawking:1998kw,Cabo-Bizet:2018ehj}
	\begin{align}
		(\tau,\phi_i)\sim(\tau+\beta,\phi_i-i\Omega_i\beta)\,,\label{eq:identification}
	\end{align}
	where $\beta=T^{-1}$ is the inverse temperature and $\Omega_i$ are the angular velocities of the horizon, both given in \eqref{eq:thermo5}.

	\item The gauge fields must have no component along the shrinking circle, which is generated by the Euclidean Killing vector $\xi=-i\xi^L=\partial_\tau-i\Omega_i\partial_{\phi_i}$ at $r=r_+$:
	\begin{align}
		\mathrm{i}_\xi A^I_{(1)}\Big|_{r=r_+}=0\,, \label{eq:regularity}
	\end{align}
	see \cite{Cabo-Bizet:2018ehj,Cassani:2019mms,Bobev:2022ocx,Bobev:2025xan} for example. This regularity condition fixes the flat connection in \eqref{eq:A5} to
	\begin{align}
		\alpha^I=-\Phi^I\,,\label{eq:flat5}
	\end{align}
	where $\Phi^I$ are the electrostatic potentials, given explicitly in \eqref{eq:Phi5}.
\end{itemize}

\medskip

\noindent {\bf{Euclidean on-shell action.}} We now set up the holographically renormalized Euclidean on-shell action. It consists of the bulk action, the Gibbons--Hawking--York (GHY) term, and the counterterms, which we introduce in turn.

The Euclidean bulk action follows from the Lorentzian one \eqref{eq:action5L} by the Wick rotation \eqref{eq:Wick}, under which differential forms are unchanged while $\star_L\to-i\star$ (Appendix~\ref{app:conventions}), together with the identification $S_{\rm bulk}=-iS^L_{\rm bulk}$:
\begin{align}
	S_{\rm bulk} &= - \frac{1}{16 \pi G_N} \int \bigg[\Big(R-\mathcal V\Big) \star 1 - \frac{1}{2} \sum_{i=1}^2  \star  d\varphi_i \wedge d\varphi_i \nn\\
	& \kern6.5em - \frac{1}{2} \sum_{I=1}^3 X_I^{-2} \star F_{(2)}^I \wedge F_{(2)}^I - \frac{i}{6} |\epsilon_{IJK}| F_{(2)}^I \wedge F_{(2)}^J \wedge A_{(1)}^K\bigg]\,, \label{eq:action5}
\end{align}
where here and below the superscript $E$ for Euclidean quantities is omitted for notational convenience while the superscript $L$ is kept for Lorentzian ones. On-shell the bulk action \eqref{eq:action5} simplifies considerably: the trace of the Einstein equation \eqref{eq:Einstein5} trades the Einstein--Hilbert and scalar kinetic terms for the potential and gauge kinetic terms, and the latter, integrated by parts with the Maxwell equations \eqref{eq:Maxwell5}, combine with the Chern--Simons term into a total derivative. The resulting bulk on-shell action reads
\begin{align}
	S_{\rm bulk} \Big|_{\rm on\text-shell} = - \frac{1}{16 \pi G_N} \int \bigg[\frac{2}{3} \mathcal V \star 1+ d\Big(\frac{1}{3} \sum_{I=1}^3 X_I^{-2} \star F_{(2)}^I \wedge A_{(1)}^I\Big)\bigg]\,. \label{eq:bulkonshell5}
\end{align}
Note that the flat connection $\alpha^Idt$ enters \eqref{eq:bulkonshell5} only through the total-derivative term.

A well-defined variational principle requires the GHY term on top of the bulk action \eqref{eq:action5}. For a $d$-dimensional Euclidean bulk with boundary at a radial cutoff, it reads
\begin{align}
	S_{\rm GHY} = - \frac{1}{8 \pi G_N} \int d^{d-1}y \sqrt{h}\, \mathcal K\,, \label{eq:GHY}
\end{align}
where $\mathcal K=h^{\mu\nu}\mathcal K_{\mu\nu}$ is the trace of the extrinsic curvature of the $(d-1)$-dimensional boundary, 
\begin{align}
	\mathcal K_{\mu \nu} = \nabla_\mu \eta_\nu - \eta_\mu \eta^\rho \nabla_\rho \eta_\nu\,, \label{eq:Kdef}
\end{align}
with $\eta^\mu$ the outward unit normal, $h_{ij}$ the induced metric, and $y^i$ the boundary coordinates.

The bulk and GHY terms diverge for asymptotically locally EAdS$_d$ backgrounds when the radial cutoff is sent to infinity. In holographic renormalization \cite{Henningson:1998gx,Balasubramanian:1999re,Emparan:1999pm,deHaro:2000vlm,Bianchi:2001kw,Skenderis:2002wp}, the divergences are cancelled by local counterterms built from the induced metric $h_{ij}$ and the scalars on the cutoff surface. For a $d$-dimensional bulk with $d\le7$ the required counterterms are \cite{Batrachenko:2004fd,Balasubramanian:1999re,Emparan:1999pm,Kraus:1999di,Bueno:2022log,Ntokos:2021duk}
\begin{align}
	S_{\rm CT} &= \sum_{\mathfrak{a}=0}^{n} S_{{\rm CT}\text{-}\mathfrak{a}}\,,  \label{eq:CT}\\
	S_{\text{CT-0}} &= \frac{1}{8 \pi G_N} \int d^{d-1}y \sqrt{h}\, \mathcal W\,, \nn\\
	S_{\text{CT-1}} &= \frac{1}{8 \pi G_N} \int d^{d-1}y \sqrt{h}\, \frac{\mathcal R}{2(d-3)g}\,, \nn\\
	S_{\text{CT-2}} &= \frac{1}{8 \pi G_N} \int d^{d-1}y \sqrt{h}\, \frac{1}{2(d-3)^2(d-5)g^3} \Big[ \mathcal R_{ij} \mathcal R^{ij} - \frac{d-1}{4(d-2)} \mathcal R^2\Big]\,,\nn\\
	S_{\text{CT-3}} &= -\frac{1}{8 \pi G_N} \int d^{d-1}y \sqrt{h}\, \frac{1}{(d-3)^3(d-5)(d-7)g^5} \bigg[ \frac{3d-1}{4(d-2)}\, \mathcal R\, \mathcal R_{ij} \mathcal R^{ij} - \frac{(d-1)(d+1)}{16(d-2)^2}\, \mathcal R^3 \nn\\
	&\quad - 2\, \mathcal R^{ij} \mathcal R_{ikjl} \mathcal R^{kl} + \frac{d-3}{2(d-2)}\, \mathcal R^{ij} \nabla_i \nabla_j \mathcal R - \mathcal R^{ij} \nabla^2 \mathcal R_{ij} + \frac{1}{2(d-2)}\, \mathcal R\, \nabla^2 \mathcal R \bigg]\,,\nn
\end{align}
where $n=\lfloor(d-1)/2\rfloor$ as introduced in \eqref{eq:mu}, $\mathcal R_{ijkl}$, $\mathcal R_{ij}$, and $\mathcal R$ are the Riemann tensor, Ricci tensor, and Ricci scalar for the boundary induced metric $h_{ij}$, and $\nabla_i$ is its covariant derivative. The leading counterterm is given by the superpotential $\mathcal W$ associated with the scalar potential, \eqref{eq:potential5} in five and \eqref{eq:potential6} in six dimensions, which reduces to the constant $(d-2)g$ of the pure-gravity counterterm \cite{Balasubramanian:1999re,Emparan:1999pm,Kraus:1999di,Bueno:2022log} when the scalars are trivial. Writing $S_{\text{CT-0}}$ with $\mathcal W$ rather than with $(d-2)g$ incorporates the finite scalar counterterms fixed by supersymmetry \cite{Freedman:2013oja,Bobev:2013cja,Bobev:2016nua,Freedman:2016yue}; for the backgrounds investigated in this paper, whose scalars fall off as $X_I-1=\mathcal O(\tir^{3-d})$, this makes a finite difference only in $d=5$, where it agrees with the counterterm of \cite{Ntokos:2021duk} for the scalar fall-off at hand, while the scalar dependence of the subleading counterterms does not contribute in any of the dimensions $d=5,6,7$ \cite{Batrachenko:2004fd}, and is therefore omitted in \eqref{eq:CT}. For odd $d$ the last counterterm $S_{\text{CT-}n}$ carries a pole $(d-2n-1)^{-1}$, which signals a logarithmic divergence; the latter is cancelled by the counterterm obtained from $S_{\text{CT-}n}$ by replacing the pole factor with $\log\varepsilon$, where $\rho=\varepsilon\simeq(g\tir_\infty)^{-2}$ is the radial cutoff in the Fefferman--Graham coordinates \eqref{eq:FG} below, and the coefficient of $\log\varepsilon$ captures the holographic Weyl anomaly \cite{Henningson:1998gx,Emparan:1999pm,deHaro:2000vlm,Bueno:2022log}. For the black holes considered in this paper, however, the conformal boundary $S^1\times S^{d-2}$ is conformally flat, so that the integrated anomaly vanishes and the logarithmic counterterm becomes trivial.  For odd $d$ we therefore drop it from the outset and let $S_{\text{CT-}n}$ denote instead the finite counterterm
\begin{align}
	S_{\text{CT-}n}\overset{\text{odd}~d}{=}\fft{\zeta}{8\pi G_N}\int d^{d-1}y\sqrt{h}\,\mR^{n}\,, \label{eq:CTfin}
\end{align}
with $\zeta$ a free constant parametrizing the scheme,\footnote{This term is not a superconformal invariant: in a supersymmetric scheme \cite{Papadimitriou:2017kzw,An:2017ihs,Ntokos:2021duk} it is absent and the parameter-independent Casimir term is retained. In this paper it serves merely as a device to remove the latter, \textit{cf.}\ \eqref{eq:scheme5}. The comparison of the Casimir term with the supersymmetric Casimir energy of the dual field theory \cite{Assel:2014paa,Assel:2015nca,BenettiGenolini:2016qwm,BenettiGenolini:2016tsn} involves instead the supersymmetry anomaly and the choice of supercurrent multiplet \cite{Papadimitriou:2017kzw,Papadimitriou:2019yug,Closset:2019ucb,Panopoulos:2023cdg,Panopoulos:2026gwt,Panopoulos:2026slv}; the local counterterm relating different multiplets is not of the form \eqref{eq:CTfin} and requires the old-minimal auxiliary fields \cite{Panopoulos:2026gwt,Panopoulos:2026slv}.} while for even $d$ the last counterterm $S_{\text{CT-}n}$ is the standard one listed in \eqref{eq:CT}.

Finally, the renormalized on-shell action of a $d$-dimensional background $\mathcal M_d$ is the sum of the three contributions evaluated at a radial cutoff, sent to infinity:
\begin{align}
	I_{\rm reg} \Big|_{\mathcal M_d} = \Big[S_{\rm bulk} + S_{\rm GHY} + S_{\rm CT}\Big]_{\mathcal M_d}\,. \label{eq:Ireg}
\end{align}

\medskip

\noindent {\bf{Asymptotically global AdS frame.}}  A bulk solution determines the metric on its conformal boundary only up to a Weyl rescaling, and in odd bulk dimension the renormalized on-shell action depends on the representative of this conformal class through the Weyl anomaly \cite{Papadimitriou:2005ii}. Holographic renormalization of a given solution therefore requires a representative to be specified; in the Fefferman--Graham (FG) gauge adapted to it \cite{Henningson:1998gx,deHaro:2000vlm,Bianchi:2001kw,Skenderis:2002wp} the metric reads
\begin{align}
	ds^2 = \frac{d\rho^2}{4g^2 \rho^2} + \frac{1}{g^2 \rho} g_{ij} (x,\rho) dx^i dx^j\,, \label{eq:FG}
\end{align}
where $g_{ij}(x,0)$ is the specified representative. The surfaces of constant $r$ of the Boyer--Lindquist-type coordinates of \eqref{eq:metric5} select the representative $\lim_{r\to\infty}r^{-2}ds^2$, which depends on the rotation parameters \cite{Gibbons:2004ai} and differs from the Einstein static universe $S^1\times S^{d-2}$ by a Weyl factor and a change of angular coordinates \cite{Gibbons:2005jd}. The dual field theory results of interest, such as the superconformal index, are formulated on the Einstein static universe, which is selected by the surfaces of constant $\tir$ in the coordinates $(\tir,\tmu_i)$ of \cite{Gibbons:2004uw}; the change of coordinates acts on $(r,\mu_i)$ only, so that the identification \eqref{eq:identification}, which encodes the temperature and the angular velocities, is unaffected. For an electrically charged rotating black hole in any dimension $d=2n+\epsilon+1$ described by the coordinates \eqref{eq:mu} the new coordinates are defined by
\begin{align}
	\Xi_i\, \tir^2 \tmu_i^2 &= \big(r^2 + a_i^2\big) \mu_i^2\,, \qquad \sum_{i=1}^{n+\epsilon} \tmu_i^2 = 1\,, \nn\\
	\to\qquad \tir^2 &= \sum_{i=1}^{n+\epsilon} \frac{r^2+a_i^2}{\Xi_i}\,\mu_i^2\,, \qquad 1+g^2\tir^2 = \big(1+g^2r^2\big) \sum_{i=1}^{n+\epsilon} \frac{\mu_i^2}{\Xi_i}\,, \label{eq:GLPP}
\end{align}
with $a_{n+1}=0$ in even $d$.  Throughout this paper, holographic renormalization is carried out for the representative selected by these coordinates. Other representatives are equally admissible, \textit{e.g.}\ the one selected by the Boyer--Lindquist radial coordinate \cite{Awad:2000aj,Papadimitriou:2005ii,Bobev:2022ocx,Awad:2022exw}; the present one allows for a direct comparison with the dual field theory results on $S^1\times S^{d-2}$ with parameter-independent Casimir term and finite counterterms as shown below.

In the new coordinates $(\tau,\tir,\tmu_i,\phi_i)$, the Euclidean saddle --- the Wick rotation \eqref{eq:Wick} of the five-dimensional black hole \eqref{eq:metric5}, or of the six- and seven-dimensional black holes of Subsection~\ref{sec:holo-renorm:AdS6} and of \cite{Bobev:2023bxl,Bobev:2025xan} --- is expanded at large $\tir$ as
\begin{align}
	ds^2 &= \bar g_{\mu\nu} dx^\mu dx^\nu + \gamma_{\mu\nu} dx^\mu dx^\nu\,, \nn\\
	\gamma_{\mu\nu} &=
	\begin{cases}
		\tir^{\,1-d} \sum_{q=0}^{\infty} \Big(\gamma^{(q)}_{\tir\tir} + \epsilon\, \hat\gamma^{(q)}_{\tir\tir}\, \tir^{\,3-d}\Big) \tir^{-2q} & (\mu\nu = \tir \tir) \\[4pt]
		\tir^{\,2-d} \sum_{q=0}^{\infty} \Big(\gamma^{(q)}_{\tir\tmu_a} + \epsilon\, \hat\gamma^{(q)}_{\tir\tmu_a}\, \tir^{\,3-d}\Big) \tir^{-2q} & (\mu\nu = \tir \tmu_a) \\[4pt]
		\tir^{\,5-d} \sum_{q=0}^{\infty} \Big(\gamma^{(q)}_{ij} + \epsilon\, \hat\gamma^{(q)}_{ij}\, \tir^{\,3-d}\Big) \tir^{-2q} & (\mu\nu = x^i x^j)\\[4pt]
		0  & (\text{otherwise})\,,
	\end{cases} \label{eq:expansion}
\end{align}
where we have used the coordinate indices $\{x^\mu\}=\{\tau,\tir,\tmu_a,\phi_i\}$ with $a=1,\dots,n+\epsilon-1$ (one direction cosine is eliminated by the constraint) along with the boundary ones $\{x^i\}=\{\tau,\tmu_a,\phi_i\}$. For odd $d$ ($\epsilon=0$) each component is expanded in integer powers of $\tir^{-2}$ relative to its leading term. For even $d$ ($\epsilon=1$) a second such series carried by the hatted coefficients appears at relative order $\tir^{3-d}$, but it contributes to none of the divergent or finite terms below. In the asymptotic expansion of a $d$-dimensional Euclidean saddle \eqref{eq:expansion} at large $\tir$, $\bar g_{\mu\nu}$ denotes the global EAdS$_d$ geometry
\begin{align}
	d\bar s^2_{{\rm EAdS}_d} &= \big(1+g^2 \tir^2 \big) d\tau^2 + \frac{d\tir^2}{1+g^2 \tir^2} + \tir^2 ds^2_{S^{d-2}}\,,\label{eq:EAdS} \\
	ds^2_{S^{d-2}} &= \sum_{i=1}^{n+\epsilon} d\tmu_i^2 + \sum_{i=1}^{n} \tmu_i^2 d\phi_i^2\,, \nn
\end{align}
and the power-law suppression of the perturbation $\gamma_{\mu\nu}$ is determined by substituting the expansions (derived from the coordinate transformation \eqref{eq:GLPP})
\begin{align}
	r^2 &= \mD\,\tir^2 - \frac{1}{\mD}\sum_{j=1}^{n+\epsilon} a_j^2 \Xi_j \tmu_j^2 + \mathcal O(\tir^{-2})\,, \nn\\
	\mu_i^2 &= \frac{\Xi_i \tmu_i^2}{\mD} \bigg[1+ \frac{1}{\tir^2} \bigg(- \frac{a_i^2}{\mD} + \frac{1}{\mD^2}\sum_{k=1}^{n+\epsilon} a_k^2 \Xi_k \tmu_k^2\bigg) + \mathcal O(\tir^{-4}) \bigg]\,, \qquad \mD \equiv \sum_{i=1}^{n+\epsilon} \Xi_i \tmu_i^2\,, \label{eq:coordinates}
\end{align}
into the Euclidean saddle of interest; in the form \eqref{eq:expansion} it is manifest that the Euclidean saddle approaches global EAdS$_d$ at large $\tir$. It is also worth mentioning that the perturbation $\gamma_{\mu\nu}$ vanishes identically for vanishing mass parameter, $m=0$, where \eqref{eq:expansion} reduces to global EAdS$_d$.

As an illustration, we rewrite the Wick-rotated five-dimensional geometry \eqref{eq:metric5} in the coordinates \eqref{eq:GLPP} as
\begin{align}
	ds_5^2 &= (H_1 H_2 H_3)^{\frac{1}{3}} \bigg[\big(1+g^2 \tir^2 \big) d\tau^2 + \frac{d\tir^2}{1+g^2 \tir^2} + \tir^2 \sum_{i=1}^{2} \Big(d \tmu_i^2 + \tmu_i^2 d\phi_i^2\Big) \nn\\
	&\kern7em + \frac{2m\mX(r)}{(1+g^2r^2)\,\Delta_r\, \tir^2 \sum_{i=1}^2 \frac{\Xi_i\tmu_i^2}{(r^2+a_i^2)^2}} \bigg(\frac{d\tir}{\tir} + \tir^2 \sum_{i=1}^2 \frac{\Xi_i\, \tmu_i d \tmu_i}{r^2+a_i^2}\bigg)^2 \nn\\
	&\kern7em + \sum_{I=1}^3\frac{2ms_I^2}{\Sigma H_I(s_I^2-s_J^2)(s_I^2-s_K^2)} K_I^2\bigg]\,, \label{eq:metric5new}
\end{align}
where the new radial coordinate reads
\begin{align}
	\tir^2 = \frac{(1+g^2r^2)\Delta_\theta}{g^2 \Xi_1 \Xi_2} - \frac{1}{g^2}\,, \label{eq:rtilde5}
\end{align}
and we have further defined
\begin{align}
	2m\mX(r) \equiv \big(1+g^2r^2\big) \big(r^2+a_1^2\big) \big(r^2+a_2^2\big)-\Delta_r(r)  \,. \label{eq:calX5}
\end{align}
Since $2m\mX(r)=\mathcal O(r^4)$, $\Delta_r=\mathcal O(r^6)$, and $K_I=\mathcal O(1)$, all $m$-dependent terms of \eqref{eq:metric5new}, which constitute the perturbation $\gamma_{\mu\nu}$, are suppressed at large $\tir$ as dictated by \eqref{eq:expansion} with $d=5$. It is straightforward to check that the six- and seven-dimensional Euclidean saddles, Wick rotated from the corresponding black holes, satisfy the asymptotic behavior of \eqref{eq:expansion} with $d=6,7$ as well.

The coordinate system $(\tau,\tir,\tmu_i,\phi_i)$ is precisely the asymptotically non-rotating frame of \cite{Bobev:2023bxl,Bobev:2025xan}, in which \cite{Bobev:2025xan} carried out the holographic renormalization of the seven-dimensional black hole with the radial cutoff at $\tir=\tir_\infty$ rather than at a constant-$r$ hypersurface. Below we implement the holographic renormalization for the five-dimensional Euclidean saddle \eqref{eq:metric5new} in the same frame, while presenting the intermediate steps for a generic bulk dimension $d$ so that its generalization to higher-dimensional black holes is straightforward.

\medskip

\noindent {\bf{Renormalized on-shell action; bulk contributions.}} We now evaluate the renormalized on-shell action \eqref{eq:Ireg} on the five-dimensional Euclidean saddle \eqref{eq:metric5new} with the radial cutoff at $\tir=\tir_\infty$, starting with the bulk contribution \eqref{eq:bulkonshell5}. The five-dimensional bulk integral is organized as follows.
\begin{itemize}
	\item The integrals over $(\tau,\phi_i)$ are trivial and give $\beta(2\pi)^n$, with $\beta=1/T$ the period of $\tau$ fixed by \eqref{eq:identification}.
	
	\item Since the horizon $r=r_+$ is a surface of constant $r$ while the cutoff is at constant $\tir$, it is convenient to use the original coordinates: at fixed $\mu_i$, $r$ runs from $r_+$ to the cutoff $r(\tir_\infty,\mu_i)$, where $r(\tir,\mu_i)$ is determined through the coordinate transformation \eqref{eq:GLPP} at fixed $\mu_i$; equivalently, at fixed $\tmu_i$, $\tir$ runs from the horizon $\tir(r_+,\tmu_i)$ to $\tir_\infty$ with $\tir(r,\tmu_i)$ defined implicitly by \eqref{eq:GLPP}. For the five-dimensional black hole, with $(\mu_1,\mu_2)=(\sin\theta,\cos\theta)$, the horizon in the new radial coordinate and the cutoff in the original one are at
	\begin{align}
		\tir(r_+,\theta)&=\sqrt{\frac{(1+g^2r_+^2)\Delta_\theta}{g^2 \Xi_1 \Xi_2} - \frac{1}{g^2}}\,, \label{eq:rtildeplus5}\\
		r(\tir_\infty,\theta)&=\frac{1}{g}\sqrt{\frac{(1+g^2\tir_\infty^2)\Xi_1\Xi_2}{\Delta_\theta}-1}\,. \label{eq:rcutoff5}
	\end{align}

	\item Finally the $\mu_i$ (or $\tmu_i$) integral is performed.
\end{itemize}
The resulting bulk contribution --- the first term in the renormalized on-shell action \eqref{eq:Ireg} for the five-dimensional Euclidean saddle $\mathcal M_5=\,$\eqref{eq:metric5new} --- is given by
\begin{align}
	S_{\rm bulk} \Big|_{\eqref{eq:metric5new}} &= \frac{\pi \beta g^2}{4G_N} \tir_\infty^4 + \frac{2\pi \beta g^2m s_A \mathcal J^{(2)}_{00}}{3G_N} \tir_\infty^2 + S_{\rm bulk}^{\rm fin}\,, \label{eq:Sbulk5}\\
	S_{\rm bulk}^{\rm fin} &= \frac{2\pi \beta m s_A \mathcal J^{(2)}_{00}}{3G_N} - \frac{\pi \beta}{12 G_N \Xi_1 \Xi_2} \Big[3g^2 \big(r_+^2+a_1^2\big)\big(r_+^2+a_2^2\big) + 4m s_A\big(1+g^2r_+^2\big)\Big] \nn\\
	&\quad-\frac{\beta}{3} \sum_{I=1}^3 \Phi^I Q_I\,, \label{eq:Sbulkfin5}
\end{align}
where $\mathcal J^{(2)}_{00}=\frac{1}{2g^2(a_1^2-a_2^2)}\log\frac{\Xi_2}{\Xi_1}$ is introduced as one of the building-block boundary integrals \eqref{eq:J5}, $\Phi^I$ are the electrostatic potentials \eqref{eq:Phi5}, and $Q_I$ are the electric charges presented in \eqref{eq:Q5}. 

In the bulk contribution \eqref{eq:Sbulk5}, the divergences of order $\tir_\infty^4$ \& $\tir_\infty^2$ and the first two finite terms of \eqref{eq:Sbulkfin5} all originate from the first term of the bulk on-shell action \eqref{eq:bulkonshell5} involving a radial integral. The total-derivative term of \eqref{eq:bulkonshell5}, on the other hand, is evaluated by Stokes' theorem: in the regular gauge \eqref{eq:flat5} the horizon does not contribute \cite{Bobev:2022ocx}, and the cutoff surface yields the last, finite term of \eqref{eq:Sbulkfin5}, since at large $\tir$ the gauge field reduces to its flat part, $A^I_{(1)}\to-\Phi^Idt$, and the remaining integrand is precisely the Gauss-law flux defining $Q_I$, so that the contribution involves the combination $\sum_I\Phi^IQ_I$.

\medskip

\noindent {\bf{Renormalized on-shell action; boundary contributions.}} The GHY term \eqref{eq:GHY} and the counterterms \eqref{eq:CT} with $d=5$ are evaluated on the cutoff surface $\tir=\tir_\infty$ in the new radial coordinate, where the induced metric can be read off from the expansion \eqref{eq:expansion} as
\begin{align}
	h_{ij}\, dy^i dy^j = ds^2\Big|_{\tir=\tir_\infty} = \big(\bar h_{ij} + \alpha_{ij}\big) dy^i dy^j\,, \quad \bar h_{ij} = \bar g_{ij}\Big|_{\tir=\tir_\infty}\,, \quad \alpha_{ij} = \gamma_{ij}\Big|_{\tir=\tir_\infty}\,, \label{eq:hinduced}
\end{align}
with $y^i=x^i$. Using the decomposition of the boundary induced metric \eqref{eq:hinduced}, we expand the necessary ingredients for the on-shell evaluation of the GHY term and the counterterms at large $\tir_\infty$ as follows. 
\begin{itemize}
	\item The outward unit normal $\eta^\mu=\ell^\mu/\sqrt{\ell^\nu\ell_\nu}$ with $\ell_\mu=\delta^{\tir}{}_\mu$ and the trace of the extrinsic curvature \eqref{eq:Kdef} are expanded in the large $\tir_\infty$ limit as
	\begin{align}
		\eta^\mu &= \sqrt{ \bar g^{\tir\tir}} \bigg[\Big(1-\underbrace{\frac{1}{2} \gamma_{\tir}{}^{\tir}}_{\mathcal O(\tir_\infty^{3-d})} +\underbrace{\frac{3}{8} \big(\gamma_{\tir}{}^{\tir}\big)^2}_{\mathcal O(\tir_\infty^{6-2d})} + \mathcal O\big(\tir_\infty^{4-2d}\big)\Big) \delta^\mu{}_{\tir} - \Big(\underbrace{\gamma_{\tir}{}^{a}}_{\mathcal O(\tir_\infty^{-d})} + \mathcal O\big(\tir_\infty^{3-2d}\big)\Big) \delta^\mu{}_a \bigg]\,, \nn\\
		\mathcal K &= \underbrace{\bar{\mathcal K}}_{\mathcal O(1)} \underbrace{- \frac{1}{2} \gamma_{\tir}{}^{\tir}\, \bar{\mathcal K} + \frac{1}{2} \sqrt{ \bar g^{\tir\tir}}\, \partial_{\tir} \alpha}_{\mathcal O(\tir_\infty^{3-d})} - \underbrace{\frac{ \sqrt{ \bar g^{\tir\tir}}}{\sqrt{\bar g} } \partial_a \Big(\sqrt{\bar g}\, \bar g^{ab} \gamma_{\tir b} \Big)}_{\mathcal O(\tir_\infty^{1-d})} \nn\\
		&\quad \underbrace{{}+ \frac{3}{8} \big(\gamma_{\tir}{}^{\tir}\big)^2 \bar{\mathcal K} - \frac{1}{4} \gamma_{\tir}{}^{\tir} \sqrt{ \bar g^{\tir\tir}}\, \partial_{\tir} \alpha - \frac{1}{4} \sqrt{ \bar g^{\tir\tir}}\, \partial_{\tir} \big(\alpha^{ij}\alpha_{ij}\big)}_{\mathcal O(\tir_\infty^{6-2d})} + \mathcal O\big(\tir_\infty^{4-2d}\big)\,, \label{eq:K}
	\end{align}
	where $\bar{\mathcal K}=\sqrt{1+g^2\tir_\infty^2}\,\big(\frac{d-2}{\tir_\infty}+\frac{g^2\tir_\infty}{1+g^2\tir_\infty^2}\big)$ is the trace of the extrinsic curvature of the cutoff surface in global EAdS$_d$, and the $d$-dimensional indices are raised with $\bar g^{\mu\nu}$ and the boundary ones with $\bar h^{ij}$; for example, $\gamma_{\tir}{}^{\tir}=\bar g^{\tir\tir}\gamma_{\tir\tir}$, $\gamma_{\tir}{}^{a}=\bar g^{ab}\gamma_{\tir b}$, $\alpha=\bar h^{ij}\alpha_{ij}$, and $\alpha^{ij}\alpha_{ij}=\bar h^{ik}\bar h^{jl}\alpha_{ij}\alpha_{kl}$, which are of order $\tir_\infty^{3-d}$, $\tir_\infty^{-d}$, $\tir_\infty^{3-d}$, and $\tir_\infty^{6-2d}$ respectively by the asymptotic expansion \eqref{eq:expansion}.
	
	\item Likewise, the volume element and the Ricci scalar of the cutoff surface are expanded around their global EAdS$_d$ values $\sqrt{\bar h}$ and $\bar{\mathcal R}=(d-2)(d-3)/\tir_\infty^{2}$ as
	\begin{align}
		\sqrt{h} &= \sqrt{\bar h}\, \bigg[1 +  \underbrace{\frac{1}{2} \alpha}_{\mathcal O(\tir_\infty^{3-d})} + \underbrace{\frac{1}{8} \alpha^2 - \frac{1}{4} \alpha^{ij}\alpha_{ij}}_{\mathcal O(\tir_\infty^{6-2d})} + \mathcal O\big(\tir_\infty^{9-3d}\big) \bigg]\,, \label{eq:sqrth} \\
		\mathcal R &= \underbrace{\bar{\mathcal R}}_{\mathcal O(\tir_\infty^{-2})} \underbrace{- \Big(\alpha^{ij} \bar {\mathcal R}_{ij} - \bar \nabla_i \bar \nabla_j \alpha^{ij} + \bar{\nabla}^2 \alpha \Big)}_{\equiv\,\delta\mathcal R\,=\,\mathcal O(\tir_\infty^{1-d})} + \mathcal O\big(\tir_\infty^{4-2d}\big)\,, \label{eq:Ricci}
	\end{align}
	where barred quantities are defined with respect to $\bar h_{ij}$.
\end{itemize}
From \eqref{eq:K} and \eqref{eq:sqrth}, the integrands $\sqrt h\,\mathcal K$ of the GHY term \eqref{eq:GHY} and $\sqrt h$ of the leading counterterm $S_{\text{CT-0}}$ in \eqref{eq:CT} are of order $\tir_\infty^{d-1}$ at zeroth order in the perturbation, of order $\tir_\infty^{2}$ at first order, and of order $\tir_\infty^{5-d}$ at second order. Hence the second-order terms survive the limit $\tir_\infty\to\infty$ only in $d=5$, whereas the first-order terms contribute in every dimension both to the divergence of order $\tir_\infty^{2}$ and, through the subleading coefficients in \eqref{eq:expansion}, at finite order; the total-derivative term of order $\tir_\infty^{1-d}$ in \eqref{eq:K}, although it is also of first order in the perturbation, integrates to zero over the closed cutoff surface. In the first subleading counterterm $S_{\text{CT-1}}$ in \eqref{eq:CT}, the first-order terms of both \eqref{eq:sqrth} and \eqref{eq:Ricci} contribute at finite order, while the second-order terms are suppressed in every dimension $d=5,6,7$. 

Now we present the boundary contributions --- the GHY term and the counterterms --- explicitly for the five-dimensional Euclidean saddle \eqref{eq:metric5new}. For the GHY contribution, we first read off from the metric \eqref{eq:metric5new} the relevant expansion coefficients of \eqref{eq:expansion} as
\begin{align}
	\gamma_{\tir \tir}^{(0)} &= - \frac{4m s_A}{3g^2 \mD}\,, \qquad \gamma_{\tir \tmu_1}^{(0)} = \frac{2m s_A \big(a_1^2-a_2^2\big)}{\mD^2}\, \tmu_1\,, \nn\\
	\gamma_{\tir \tir}^{(1)} &= \frac{2m}{g^4 \mD^2} \Big[1 + 2s_A + g^2\big(a_1^2+a_2^2\big) s_A + g^2\big(a_1^2+a_2^2+g^2a_1^2a_2^2\big) s_B + 4g^4a_1^2a_2^2\, s_C\Big] \nn\\
	&\quad + \frac{4m^2\big(5s_A^2-6s_B\big)}{9g^2\mD^2} - \frac{2ms_A}{3g^4\mD} - \frac{2ms_A}{3g^2\mD^3} \Big[\big(5a_1^2+a_2^2\big)\Xi_1 \tmu_1^2 + \big(a_1^2+5a_2^2\big)\Xi_2 \tmu_2^2\Big]\,, \label{eq:gamma5}
\end{align}
together with the trace $\alpha=\alpha^{(-2)}\tir_\infty^{-2}+\alpha^{(-4)}\tir_\infty^{-4}+\mathcal O(\tir_\infty^{-6})$ of the perturbation $\alpha_{ij}$ in the induced metric \eqref{eq:hinduced}, whose expansion coefficients are given by
\begin{align}
	\alpha^{(-2)} &= \frac{8ms_A}{3 \mD}\,, \nn\\
	\alpha^{(-4)} &= - \frac{8ms_A}{3\mD^2}\frac{a_1^2-a_2^2}{\mD} \Big(\Xi_2 \tmu_2^2 -\Xi_1 \tmu_1^2\Big) - \frac{2mg^2 s_A \big(a_1^2-a_2^2\big)^2}{\mD^3}\tmu_1^2 \tmu_2^2 \nn\\
	&\quad - \frac{16m^2 \big(s_A^2-3s_B\big)}{9\mD^2} + \frac{2m}{\mD^3} \sum_{I=1}^3 \frac{s_I^2}{\prod_{J\neq I}(s_I^2-s_J^2)}\Big(- \frac{\kappa_I^2}{g^2} + \sum_{i=1}^2 \kappa_{Ii}^2\tmu_i^2\Big) \,, \label{eq:alpha5}
\end{align}
where $\kappa_I$ and $\kappa_{Ii}$ are the constants introduced in \eqref{eq:oneforms5}. Note that, since the leading perturbation $\gamma^{(0)}_{ij}$ of the boundary metric, $\alpha_{ij}=\gamma^{(0)}_{ij}+\mathcal O(\tir_\infty^{-2})$, is proportional to $\bar h_{ij}$ for the saddle \eqref{eq:metric5new}, one has $\alpha^{ij}\alpha_{ij}=\frac14\alpha^2+\mathcal O(\tir_\infty^{-6})$, so that the volume element \eqref{eq:sqrth} can be simplified further for $d=5$ as
\begin{align}
	\sqrt{h} = \sqrt{\bar h} \bigg[1+  \frac{\alpha^{(-2)}}{2\tir_\infty^2} + \frac{1}{\tir_\infty^4} \bigg(\frac{\alpha^{(-4)}}{2} +\frac{\big(\alpha^{(-2)}\big)^2}{16}\bigg)+ \mathcal O \big(\tir_\infty^{-6}\big)\bigg]\,. \label{eq:sqrth5}
\end{align}
Substituting \eqref{eq:gamma5} and \eqref{eq:alpha5} into \eqref{eq:K} and \eqref{eq:sqrth5}, and performing the boundary integrals (Appendix~\ref{app:int5}) gives the GHY contribution \eqref{eq:GHY} for the five-dimensional saddle \eqref{eq:metric5new},
\begin{align}
	S_{\rm GHY} \Big|_{\eqref{eq:metric5new}}  &= -\frac{\pi \beta g^2}{G_N} \tir_{\infty}^4 - \frac{\pi \beta}{6G_N} \bigg(\frac{9}{2} + 16 g^2 m s_A \mathcal J^{(2)}_{00}\bigg)\tir_{\infty}^2+S_{\rm GHY}^{\rm fin}\,, \label{eq:SGHY5}\\
	S_{\rm GHY}^{\rm fin} &= \frac{\pi \beta m}{3G_N} \bigg[\frac{\mathcal F}{\Xi_1 \Xi_2} - 11 s_A \mathcal J^{(2)}_{00}\bigg]\,, \nn\\
	\mathcal F &\equiv 3\big(1+2s_A\big) + \Big[3g^2\big(a_1^2+a_2^2+g^2a_1^2a_2^2\big) - 4g^2 m\Big] s_B + 12 g^4 a_1^2 a_2^2 s_C\,. \label{eq:SGHYfin5}
\end{align}

\medskip

The counterterms \eqref{eq:CT} with $d=5$ are evaluated in the same way. In $S_{\text{CT-0}}$ both the terms of order $\tir_\infty^{-4}$ in \eqref{eq:sqrth5} and the term of order $\tir_\infty^{-4}$ in $\mathcal W-3g=g\sum_I(X_I-1)$ --- read off from the scalars \eqref{eq:X5} by employing the coordinate expansion \eqref{eq:coordinates}, the terms of order $\tir_\infty^{-2}$ cancelling in the sum over $I$ --- contribute at finite order, while in $S_{\text{CT-1}}$ and $S_{\text{CT-2}}$ the first and the zeroth order suffice, respectively. Altogether, the counterterms evaluate on \eqref{eq:metric5new} to

\begin{align}
	S_{\text{CT-0}} \Big|_{\eqref{eq:metric5new}}  &=  \frac{3g^2 \pi \beta}{4G_N} \tir_{\infty}^4 + \frac{\pi \beta}{8G_N} \big(3 + 16 g^2 m s_A \mathcal J^{(2)}_{00}\big)\tir_{\infty}^2+S_{\text{CT-0}}^{\rm fin}\,, \nn\\
	S_{\text{CT-1}} \Big|_{\eqref{eq:metric5new}}  &=   \frac{3 \pi \beta}{8G_N} \tir_{\infty}^2 + \frac{\pi \beta}{16g^2G_N} \big(3 + 8 g^2 m s_A \mathcal J^{(2)}_{00}\big)\,, \nn\\
	S_{\text{CT-2}} \Big|_{\eqref{eq:metric5new}}  &=   \frac{9 \pi \beta g\, \zeta}{G_N}\,, \label{eq:SCT5}
\end{align}
where 
\begin{align}
	S_{\text{CT-0}}^{\rm fin} = - \frac{\pi \beta}{32G_N g^2} \bigg[3 + \frac{8g^2m\,\mathcal F}{\Xi_1 \Xi_2} - 80 g^2 m s_A \mathcal J^{(2)}_{00}\bigg]\,. \label{eq:SCT0fin5}
\end{align}

\medskip

\noindent {\bf{Renormalized on-shell action; total.}} Adding \eqref{eq:Sbulk5}, \eqref{eq:SGHY5}, and \eqref{eq:SCT5}, the divergences of order $\tir_\infty^4$ and $\tir_\infty^2$ cancel and the renormalized on-shell action \eqref{eq:Ireg} for the five-dimensional Euclidean saddle \eqref{eq:metric5new} is given by
\begin{align}
	I_{\rm reg}^{(5)}&\equiv I_{\rm reg}\Big|_{\eqref{eq:metric5new}} \nn\\
	&= S_{\rm bulk}^{\rm fin} +S_{\rm GHY}^{\rm fin}+ S_{\text{CT-0}}^{\rm fin} + \frac{\pi \beta}{16g^2G_N} \big(3 + 8 g^2 m s_A \mathcal J^{(2)}_{00}\big) + \frac{9 \pi \beta g\,\zeta}{G_N} \nn\\
	&= \beta \mathcal I^{(5)}\,, \label{eq:Ireg5}
\end{align}
where the finite quantity $\mathcal I^{(5)}$ reads
\begin{align}
	\mathcal I^{(5)} &= \frac{3\pi}{32g^2 G_N} + \frac{\pi}{12 G_N \Xi_1 \Xi_2} \Big[m\,\mathcal F - 3g^2 \big(r_+^2+a_1^2\big)\big(r_+^2+a_2^2\big) - 4m s_A\big(1+g^2r_+^2\big)\Big] \nn\\
	&\quad - \frac{1}{3} \sum_{I=1}^3 \Phi^I Q_I + \frac{9 \pi g\,\zeta}{G_N}\,, \label{eq:I5}
\end{align}
with $\Phi^I$ and $Q_I$ given in \eqref{eq:Phi5} and \eqref{eq:Q5} respectively.

\medskip

\noindent {\bf{Thermodynamics and QSR.}} Now we review the thermodynamics of the five-dimensional black hole \eqref{eq:metric5} following \cite{Wu:2011gq} with the Newton constant $G_N$ restored. The entropy, temperature, and horizon angular velocities are given by 
\begin{align}
	S &= \frac{\pi^2 \sqrt{W}}{2G_N \Xi_1 \Xi_2 r_+}\,, \qquad T = \frac{1}{\beta} = \frac{\Delta_r^\prime(r_+)}{4\pi \sqrt{W}}\,, \nn\\
	\Omega_i &= \frac{2mr_+^2}{W} \bigg\{a_i c_1 c_2 c_3 \sqrt{\Xi_{1i} \Xi_{2i} \Xi_{3i}} \Big[r_+^2+ a_j^2 -2mg^2 a_j^2 \big(s_B + 2g^2 a_j^2 s_C\big)\Big]  \label{eq:thermo5}\\
	&\kern5em - a_j s_1 s_2 s_3 \sqrt{\Xi_{1j} \Xi_{2j} \Xi_{3j}} \Big[\big(r_+^2+a_j^2\big)\Xi_i^2 \nn\\
	&\kern9em -2m\big(1+g^2a_i^2+2g^2a_i^2 s_A+g^2a_i^2 \big(1+g^2a_i^2\big) s_B +2g^4a_i^4 s_C\big)\Big]\bigg\}\,,\nn
\end{align}
where, here and below, $j\neq i$ denotes the other rotation index. The electrostatic potentials $\Phi^I=(\xi^L)^\mu A^I_\mu\big|_{r=r_+}-(\xi^L)^\mu A^I_\mu\big|_{r\to\infty}$ with $\xi^L=\partial_t+\Omega_i\partial_{\phi_i}$ (see \cite{Cvetic:2005zi,Chen:2005zj,Cabo-Bizet:2018ehj,Cassani:2019mms} for example) and the function $W$ appearing in \eqref{eq:thermo5} read 
\begin{align}
	\Phi^I &= \frac{2mr_+^2}{W} \bigg\{ \frac{c_I^2}{c_1 c_2 c_3} P_I \Big[\big(r_+^2+a_1^2\big)\big(r_+^2+a_2^2\big) + 2m \big(s_A-s_I^2\big) \big(r_+^2 - g^2 a_1^2 a_2^2 s_I^2\big) \nn\\
	&\kern9em - 4mg^2 a_1^2 a_2^2 s_C - 2m \big(a_1^2+a_2^2-2m\big) \frac{s_C}{s_I^2}\Big] \nn\\
	&\kern5em - \frac{a_1 a_2\, c_1 c_2 c_3}{c_I^2} N_I \Big[g^2\big(r_+^2+a_1^2\big)\big(r_+^2+a_2^2\big) - 2m \big(1+2s_I^2\big) \nn\\
	&\kern13em + 2mg^2 \big(s_A-s_I^2\big) \big(r_+^2 - s_I^2 (a_1^2+a_2^2)\big) \nn\\
	&\kern13em +2mg^2 \big(2m-(1+2s_I^2)g^2 a_1^2a_2^2\big) \frac{s_C}{s_I^2}\Big] \bigg\}\,, \label{eq:Phi5}\\
	W &= \Big[\big(r_+^2+a_1^2\big)\big(r_+^2+a_2^2\big)+2mr_+^2 s_A\Big] \Big[\big(r_+^2+a_1^2\big)\big(r_+^2+a_2^2\big) -4mg^2 a_1^2 a_2^2 s_B \nn\\
	&\kern11em + 2mg^2 \big((a_1+a_2)^2-g^2 a_1^2 a_2^2\big) \big((a_1-a_2)^2-g^2 a_1^2 a_2^2\big) s_C \Big] \nn\\
	&\quad +8m^2r_+^2 a_1 a_2\, c_1 c_2 c_3 s_1 s_2 s_3 \sqrt{\Xi_{11} \Xi_{21} \Xi_{31} \Xi_{12} \Xi_{22} \Xi_{32}} +4m^2r_+^2 \big(r_+^2 +g^2 a_1^2 a_2^2 s_A\big) s_B \nn\\
	&\quad -4m^2 \Big[\big(a_1^2+a_2^2\big) \big(1+g^2 a_1^2\big)\big(1+g^2a_2^2\big) r_+^2 + g^2\big((a_1^4 +a_2^4)r_+^2+g^2 a_1^4 a_2^4 (2+g^2 r_+^2)\big) s_A \nn\\
	&\kern5em +g^2 a_1^2 a_2^2 \big(a_1^2+a_2^2 + g^2 a_1^2 a_2^2\big) \big(2+g^2r_+^2 s_B\big)\Big] s_C \nn\\
	&\quad + 4m^2 g^4 a_1^4 a_2^4 \big(s_B^2-2s_As_C\big) - 8m^2r_+^2 g^6 a_1^4 a_2^4 s_C^2+8m^3 \big(r_+^2+g^2a_1^2 a_2^2\big) s_C\,. \label{eq:calW5}
\end{align}
The energy, angular momenta, and electric charges are
\begin{align}
	E &= \frac{\pi m}{4G_N \Xi_1 \Xi_2} \bigg[2c_1^2 c_2^2 c_3^2 \Big(\frac{1}{\Xi_1}+ \frac{1}{\Xi_2}-1\Big)+1-s_B \big(1+\Xi_1 \Xi_2\big)\nn\\
	&\kern6em-2s_C \Big(1+\big(2-\Xi_1-\Xi_2\big) \Xi_1 \Xi_2\Big)\bigg]\,, \\
	J_i &= \frac{\pi m}{2G_N \Xi_i^2 \Xi_j} \Big(a_i c_1 c_2 c_3 \sqrt{\Xi_{1i} \Xi_{2i} \Xi_{3i}} - a_j \Xi_i^2 s_1 s_2 s_3 \sqrt{\Xi_{1j} \Xi_{2j} \Xi_{3j}}\Big)\,, \label{eq:charges5}\\
	Q_I &= \frac{\pi m}{2 G_N \Xi_1 \Xi_2}\, \kappa_I\,, \label{eq:Q5}
\end{align}
with $\kappa_I$ presented in \eqref{eq:oneforms5}. 

The above thermodynamic quantities satisfy the first law of black hole thermodynamics
\begin{align}
	dE = T\, dS + \sum_{i=1}^2 \Omega_i\, dJ_i + \sum_{I=1}^3 \Phi^I dQ_I\,, \label{eq:firstlaw5}
\end{align}
over the six-dimensional parameter space $(m,a_i,\delta_I)$. Moreover, the QSR expresses the on-shell action of the black hole in terms of these thermodynamic quantities as
\begin{align}
	I_{\rm QSR}^{(5)} = \beta E - S - \beta \sum_{i=1}^2 \Omega_i J_i  - \beta \sum_{I=1}^3  \Phi^I Q_I\,, \label{eq:IQSR5}
\end{align}
where the subscript distinguishes the on-shell action predicted by black hole thermodynamics from the holographically renormalized one \eqref{eq:Ireg5}.

In the minimal scheme $\zeta=0$, the holographically renormalized on-shell action \eqref{eq:Ireg5} is related to the on-shell action determined via the QSR \eqref{eq:IQSR5} by
\begin{align}
	I_{\rm reg}^{(5)}\Big|_{\zeta=0} = \beta E_{{\rm AdS}_5} + I_{\rm QSR}^{(5)}\,, \label{eq:IregQSR5}
\end{align}
where 
\begin{align}
	E_{{\rm AdS}_5} = \frac{3\pi}{32g^2G_N}  \label{eq:Casimir5}
\end{align}
is the Casimir energy of global AdS$_5$ \cite{Balasubramanian:1999re,Papadimitriou:2005ii,Bobev:2022ocx}: $\beta E_{{\rm AdS}_5}$ is the on-shell action of global EAdS$_5$ evaluated via holographic renormalization in the same minimal scheme.  Alternatively, the scheme $\zeta=-1/(96g^3)$ removes the Casimir term for all values of the parameters, so that the two expressions coincide:
\begin{align}
	I_{\rm reg}^{(5)}\Big|_{\zeta=-1/(96g^3)} = I_{\rm QSR}^{(5)}\,. \label{eq:scheme5}
\end{align}
It is worth mentioning that the relation \eqref{eq:IregQSR5} is in accordance with the general result that the renormalized on-shell action satisfies the QSR with the holographic charges \cite{Papadimitriou:2005ii}: the holographic energy defined through the boundary stress tensor \cite{Balasubramanian:1999re,deHaro:2000vlm,Skenderis:2000in} includes the Casimir energy $E_{{\rm AdS}_5}$, whereas the energy $E$ above vanishes for global AdS$_5$.

In summary, \eqref{eq:IregQSR5} and \eqref{eq:scheme5} show that the holographically renormalized on-shell action of the general AdS$_5$ black hole \eqref{eq:metric5}, evaluated in the asymptotically global AdS frame, satisfies the QSR \eqref{eq:IQSR5} up to the scheme-dependent Casimir term of global AdS$_5$, which is removed by the finite counterterm $S_{\text{CT-2}}$ in \eqref{eq:CTfin} with the choice $\zeta=-1/(96g^3)$.  To the best of our knowledge, \eqref{eq:Ireg5}--\eqref{eq:I5} is the first explicit evaluation of the holographically renormalized on-shell action of the general rotating electrically charged AdS$_5$ black hole of the STU model; holographic renormalization itself was established for the STU model in general \cite{Ntokos:2021duk}, with the on-shell action evaluated explicitly for the solutions with $\mathrm{SU}(2)\times\mathrm{U}(1)$ isometry, \textit{i.e.}\ with equal rotation parameters. We close this subsection with two aspects of this result that deserve emphasis. 
\begin{itemize}
	\item First, the asymptotically global AdS frame of \cite{Bobev:2025xan}, employed there for the seven-dimensional black hole, works equally well for the five-dimensional STU black hole: the induced metric on the cutoff surface approaches the Einstein static universe $S^1\times S^3$ for all values of the parameters, so that the Casimir term \eqref{eq:Casimir5} and the on-shell value of the finite counterterm $S_{\text{CT-2}}$ are universal constants and a single, parameter-independent choice of $\zeta$ removes the former for the entire family. This is not the case for the representative selected by the Boyer--Lindquist radial coordinate, $\lim_{r\to\infty}r^{-2}ds^2$. Written in the angular coordinates $\tmu_i$ of \eqref{eq:coordinates} instead of $\mu_i$, it is the Einstein static universe multiplied by $\mD^{-1}$, a Weyl factor that depends on the rotation parameters; the Casimir term and the on-shell value of the finite counterterm then depend on the rotation parameters as well, with different functional forms \cite{Bobev:2022ocx}, so that no parameter-independent $\zeta$ removes the former.\footnote{For equal rotation parameters, $\tir$ becomes a function of $r$ alone, so that the constant-$r$ and constant-$\tir$ cutoff surfaces coincide; this is the case for the equal-rotation black hole whose on-shell action was computed in \cite{Cassani:2019mms,Ntokos:2021duk}.} This is in line with \cite{Gibbons:2005jd}, where the Casimir energy of Kerr--AdS was found to be independent of the rotation parameters for the Einstein static universe as the boundary metric but not for other choices, and to agree, in the minimal scheme $\zeta=0$, with the Casimir energy of free $\mathcal N=4$ SYM on $S^1\times S^3$ \cite{Balasubramanian:1999re,Gibbons:2005jd}.
	
	\item Second, within this frame, the five-dimensional STU black hole involves two ingredients that drop out for the seven-dimensional black hole of \cite{Bobev:2025xan}: the scalars of the two vector multiplets, whose fall-off $X_I-1=\mathcal O(\tir^{-2})$ lets the superpotential counterterm $S_{\text{CT-0}}$ contribute at finite order for unequal charges (in six and seven dimensions the scalars fall off as $\tir^{-3}$ and $\tir^{-4}$, and no such finite term survives \cite{Batrachenko:2004fd}), and the second-order terms of the boundary expansions \eqref{eq:K} and \eqref{eq:sqrth}, which survive the limit $\tir_\infty\to\infty$ only in $d=5$.

\end{itemize}
%

\subsection{AdS$_6$ black hole in $\mN=4$ $\SU(2)$ gauged supergravity}\label{sec:holo-renorm:AdS6}

\noindent {\bf{Theory.}} The six-dimensional theory of interest is the $\mathcal N=4$ $\mathrm{SU}(2)$ gauged supergravity of Romans \cite{Romans:1985tw}, the half-maximal gauged supergravity in six dimensions, which arises as a consistent truncation of massive type IIA supergravity on (a hemisphere of) $S^4$ \cite{Cvetic:1999un}, in the $\mathrm{U}(1)$ truncation of \cite{Chow:2008ip} that retains the metric, the dilaton, the two-form potential, and a single $\mathrm{U}(1)\subset\mathrm{SU}(2)$ gauge field; the black hole we study is the most general known black hole of this theory, with two independent angular momenta and one electric charge \cite{Chow:2008ip}. It is convenient, also for later use, to present the computation in the $\mathrm{U}(1)\times\mathrm{U}(1)$ gauged supergravity of \cite{Chow:2011fh}, whose bosonic fields consist of the metric, two abelian gauge fields $A^I_{(1)}$, a two-form potential $A_{(2)}$, and two scalars, and which reduces to the former in the sector with $X_1=X_2$ and $A^1_{(1)}=A^2_{(1)}$. This theory is the $\mathrm{U}(1)\times\mathrm{U}(1)$-invariant truncation of the matter-coupled $F(4)$ gauged supergravity \cite{DAuria:2000afl,Andrianopoli:2001rs} with one vector multiplet, in which the Cartan $\mathrm{U}(1)\subset\mathrm{SU}(2)$ and the $\mathrm{U}(1)$ of the vector multiplet are retained \cite{Karndumri:2015eta,Suh:2018szn,Couzens:2022lvg}, and it is embedded consistently in massive type IIA supergravity on the same hemisphere of $S^4$, including the two-form potential \cite{Couzens:2022lvg} (for $A_{(2)}=0$ see \cite{Cvetic:1999xx,Faedo:2021nub}). We work directly in Euclidean signature; with the conventions of Appendix~\ref{app:conventions}, the bosonic action reads\footnote{The Hodge star operator follows the convention of Appendix~\ref{app:conventions}, with the orientation of the totally anti-symmetric symbol fixed by $\epsilon_{\tau r y_1 y_2 \phi_1 \phi_2}=\sqrt{g}$ in the coordinates of \eqref{eq:metric6}; as in Subsection~\ref{sec:holo-renorm:AdS5}, this fixes, for the solution below, the sign of the Chern--Simons terms relative to the Hodge-dual terms. The Jacobian determinant of the coordinate change \cite[(2.7)--(2.8)]{Chow:2008ip} from the Jacobi--Carter coordinates $(t',\psi_1,\psi_2)$ to ours is negative for the ranges \eqref{eq:ranges6}, namely $\partial(t',\psi_1,\psi_2)/\partial(t,\phi_1,\phi_2)=-\big[\Xi_1\Xi_2\,a_1a_2\,(a_2^2-a_1^2)\big]^{-1}$, so that the orientation natural to those coordinates is the opposite of ours; accordingly, the Chern--Simons terms of \eqref{eq:action6} carry the opposite overall sign to those of \cite{Chow:2008ip}, Wick rotated as in Subsection~\ref{sec:holo-renorm:AdS5}, while the solution below, in particular the two-form potential, is that of \cite{Chow:2008ip} as it stands.}
\begin{align}
	S_{\rm bulk} &= - \frac{1}{16 \pi G_N} \int \bigg[\Big(R-\mathcal V\Big) \star 1 - \frac{1}{2} \sum_{I=1}^2  \star  d\varphi_I \wedge d\varphi_I - \frac{1}{2} \sum_{I=1}^2 X_I^{-2} \star F_{(2)}^I \wedge F_{(2)}^I \nn\\
	& \kern7em - \frac{1}{2} (X_1 X_2)^2 \star F_{(3)} \wedge F_{(3)} -g^2 (X_1 X_2)^{-1} \star A_{(2)} \wedge A_{(2)} \nn\\
	& \kern7em + i F_{(2)}^1 \wedge F_{(2)}^2 \wedge A_{(2)} + i \frac{g^2}{3} A_{(2)} \wedge A_{(2)} \wedge A_{(2)}\bigg]\,. \label{eq:action6}
\end{align}
The scalars are parametrized as
\begin{align}
	X_1 = e^{-\frac{\varphi_1}{\sqrt 8} - \frac{\varphi_2}{\sqrt 2} }  \,,\qquad  X_2 = e^{-\frac{\varphi_1}{\sqrt 8} + \frac{\varphi_2}{\sqrt 2} }\,, \label{eq:param6}
\end{align}
the field strengths are  
\begin{align}
	F_{(2)}^I = dA_{(1)}^I \,,\qquad F_{(3)} = dA_{(2)}\,, \label{eq:F6}
\end{align}
and the scalar potential is given in terms of the superpotential $\mathcal W$ of the $\mathrm{U}(1)\times\mathrm{U}(1)$ theory as
\begin{align}
	\mathcal V &= -g^2 \Big(9 X_1 X_2 + 6 X_1^{-3/2} X_2^{-1/2} +  6 X_1^{-1/2} X_2^{-3/2} - (X_1 X_2)^{-3} \Big) \nn\\
	&= 2\sum_{I=1}^2 \big(\partial_{\varphi_I}\mathcal W\big)^2 - \frac{5}{4}\,\mathcal W^2\,, \label{eq:potential6}\\
	\mathcal W &= g\bigg[\frac{3}{2}\big(X_1+X_2\big) + \big(X_1X_2\big)^{-3/2}\bigg]\,, \nn
\end{align}
where $g$ is the gauge coupling. At $X_1=X_2=X$ the superpotential reduces to that of Romans' theory, $\mathcal W=g\big(3X+X^{-3}\big)$ \cite{Batrachenko:2004fd,Suh:2018qyv}, and it is stationary at the supersymmetric AdS$_6$ vacuum $X_I=1$ \cite{Skenderis:1999mm}, where $\mathcal W=4g$ and $\mathcal V=-20g^2$.

The Einstein equation derived from \eqref{eq:action6} is
\begin{align}
	R_{\mu \nu} - \frac{1}{2} g_{\mu \nu} \big(R - \mathcal V\big) &= \sum_{I=1}^2  \frac{1}{2} \bigg[ \partial_\mu \varphi_I \partial_\nu \varphi_I - \frac{1}{2} g_{\mu \nu} g^{\rho \sigma} \partial_\rho \varphi_I \partial_\sigma \varphi_I \bigg] \nn\\
	&\quad + \sum_{I=1}^2  \frac{X_I^{-2}}{2}  \bigg[ F^{I}_{\mu}{}^{\lambda} F^{I}_{\nu \lambda}  - \frac{1}{4} g_{\mu \nu} F_{\rho \sigma}^I F^{I\rho \sigma}\bigg] \nn\\
	&\quad + \frac{(X_1 X_2)^2}{4}  \bigg[ F_{\mu}{}^{\lambda \rho} F_{\nu \lambda \rho}  - \frac{1}{6} g_{\mu \nu} F_{\rho \sigma \lambda} F^{\rho \sigma \lambda}\bigg] \nn\\
	&\quad + g^2 (X_1 X_2)^{-1} \bigg[A_{\mu}{}^{\rho} A_{\nu \rho} - \frac{1}{4} g_{\mu \nu} A_{\rho \sigma} A^{\rho \sigma} \bigg]\,. \label{eq:Einstein6}
\end{align}
The scalar field equations are given by
\begin{align}
	\Box \varphi_1 &= \frac{1}{4\sqrt 2} \sum_{I=1}^2 X_I^{-2}  F_{\mu \nu}^{ I} F^{I\mu \nu}  - \frac{1}{6\sqrt 2} (X_1 X_2)^2  F_{\mu \nu \rho} F^{\mu \nu \rho} + \frac{g^2}{2\sqrt 2} (X_1 X_2)^{-1} A_{\mu \nu} A^{\mu \nu} \nn\\
	&\quad + \frac{3g^2}{\sqrt 2} \Big(3X_1 X_2 -2X_1^{-3/2} X_2^{-1/2} - 2X_1^{-1/2} X_2^{-3/2} +(X_1 X_2)^{-3} \Big)\,, \nn\\
	\Box \varphi_2 &=  \frac{1}{2\sqrt 2} \Big(X_1^{-2} F_{\mu \nu}^{1} F^{1\mu \nu} - X_2^{-2} F_{\mu \nu}^{2} F^{2\mu \nu}\Big) + 3\sqrt{2} g^2 \Big(X_1^{-1/2} X_2^{-3/2} -X_1^{-3/2} X_2^{-1/2}\Big)\,, \label{eq:scalarEOM6}
\end{align}
and the equations of motion of the one-form and two-form potentials are
\begin{align}
	d\Big(X_1^{-2} \star F_{(2)}^1 \Big) - i F_{(2)}^2 \wedge F_{(3)} &= 0\,, \nn\\
	d\Big(X_2^{-2} \star F_{(2)}^2 \Big) - i F_{(2)}^1 \wedge F_{(3)} &= 0\,, \nn\\
	d\Big(X_1^2 X_2^2 \star F_{(3)} \Big) - i F_{(2)}^1 \wedge F_{(2)}^2 + 2g^2 (X_1 X_2)^{-1} \star A_{(2)}  - ig^2  A_{(2)} \wedge A_{(2)}  &= 0\,. \label{eq:formEOM6}
\end{align}

\medskip

\noindent {\bf{Euclidean saddle.}} The black hole of \cite{Chow:2008ip}, embedded in the $\mathrm{U}(1)\times\mathrm{U}(1)$ theory through $X_1=X_2$ and $A^1_{(1)}=A^2_{(1)}$, carries two independent rotation parameters and two equal electric charges. We present its Euclidean section directly, in the coordinates $(\tau,r,y_1,y_2,\phi_1,\phi_2)$ and with the global EAdS$_6$ part separated from the mass-dependent terms:
\begin{align}
	ds_6^2 &= H^{\frac{1}{2}} \bigg[\big(1+g^2r^2\big) \sum_{i=1}^3 \frac{\mu_i^2}{\Xi_i} d\tau^2 + \frac{U(r)}{V(r)-2m\Upsilon(r)} dr^2 +  \sum_{i=1}^3 \frac{r^2 + a_i^2}{\Xi_i} d\mu_i^2 +\sum_{i=1}^2 \frac{r^2 + a_i^2}{\Xi_i} \mu_i^2 d\phi_i^2 \nn\\
	&\kern3em -\frac{g^2}{ (1+g^2r^2) \sum_{i=1}^3 \frac{\mu_i^2}{\Xi_i} } \Big(\sum_{j=1}^3 \frac{r^2 + a_j^2}{\Xi_j} \mu_j d\mu_j \Big)^2 \nn\\
	&\kern3em +  \sum_{i=1}^2 \frac{Y_i}{\big(r^2+y_i^2\big)\big(y_i^2-y_j^2\big)} \Big(\frac{H-1}{H} \mathcal A - 2 \mathcal A_i\Big)\frac{H-1}{H} \mathcal A \nn\\
	&\kern3em + \frac{H^2\big(1+g^2r^2\big) \big(r^2+a_1^2\big)\big(r^2+a_2^2\big)- r\big(V(r)-2m\Upsilon(r)\big)}{H^2 \big(r^2+y_1^2\big)\big(r^2+y_2^2\big)} \mathcal A^2\bigg]\,, \label{eq:metric6}
\end{align}
where the direction cosines $\mu_{1,2,3}$ are those of \eqref{eq:mu} with $(n,\epsilon)=(2,1)$ and $a_3=0$ and we have also defined
\begin{align}
	H &= 1+ \frac{2m s^{ 2} r}{\big(r^2+y_1^2\big)\big(r^2+y_2^2\big)}\,, \qquad s = \sinh\delta\,, \qquad c = \cosh\delta\,, \nn\\
	\mu_i &= \sqrt{\frac{\big(a_i^2-y_1^2\big)\big(a_i^2-y_2^2\big)}{a_i^2 \big(a_i^2-a_j^2\big)} }\,, \qquad \mu_3 = \frac{y_1 y_2}{a_1 a_2}\,, \qquad \Xi_i = 1-a_i^2g^2\,,  \nn\\
	Y_i &= -\big(1-g^2y_i^2\big) \big(a_1^2-y_i^2\big) \big(a_2^2-y_i^2\big)\,, \nn\\
	U(r) &= r\, \sum_{i=1}^3  \frac{\mu_i^2}{r^2+a_i^2} \prod_{j=1}^2 \big(r^2+a_j^2\big)\,, \qquad V(r) = \frac{1}{r} \big(1+g^2r^2\big) \prod_{j=1}^2 \big(r^2+a_j^2\big)\,, \nn\\
	\Upsilon(r) &= 1- g^2 s^{2} \big(a_1^2+a_2^2\big) -2g^2r^2 s^2 - \frac{2g^2 m s^4}{r}\,, \nn\\
	\mathcal A &= - i\sum_{j=1}^3 \frac{\mu_j^2}{\Xi_j} d\tau - \sum_{j=1}^2 \frac{a_j \mu_j^2}{\Xi_j} d\phi_j\,, \nn\\
	\mathcal A_i &= - i\,\frac{1+g^2r^2}{1-g^2y_i^2} \sum_{j=1}^3 \frac{\mu_j^2}{\Xi_j} d \tau - \sum_{j=1}^2 \frac{r^2+a_j^2}{a_j^2-y_i^2} \frac{a_j \mu_j^2}{\Xi_j} d\phi_j\,. \label{eq:metric6def}
\end{align}
Here and below, a free index $j$ denotes the rotation index other than $i$, namely $j\in\{1,2\}\setminus\{i\}$, as in Subsection~\ref{sec:holo-renorm:AdS5}. The two scalars are identified as
\begin{align}
	X_1 = X_2 \equiv X = H^{-\frac{1}{4}}\,. \label{eq:X6}
\end{align}
The gauge fields, whose flat part $i\Phi\,d\tau$ is fixed by the regularity condition \eqref{eq:regularity} as in Subsection~\ref{sec:holo-renorm:AdS5}, are
\begin{align}
	A^1_{(1)} = A^2_{(1)} \equiv A_{(1)} = \frac{H-1}{H} \frac{c}{s}\, \mathcal A + i\Phi\, d\tau\,, \label{eq:A6}
\end{align}
where
\begin{align}
	\Phi = \frac{2m r_+ s c}{D_+}\,, \qquad D_+ \equiv \big(r_+^2 +a_1^2\big)\big(r_+^2 +a_2^2\big) + 2m r_+ s^{2}\,, \label{eq:Phi6}
\end{align}
is the electrostatic potential \cite{Chow:2008ip} written in terms of the outer horizon radius $r_+$, the largest real root of $V(r)-2m\Upsilon(r)=0$. The two-form potential is 
\begin{align}
	A_{(2)} &= \frac{2m s^{ 2}}{H \big(r^2+y_1^2\big)^2 \big(r^2+y_2^2\big)^2}  \bigg[ - \frac{y_1 y_2 \big(2r \big(2r^2+y_1^2+y_2^2\big)+2m s^{2}\big)}{H}  dr \wedge \mathcal A \label{eq:A2form6}\\
	&\quad + \sum_{i=1}^2 y_j \Big(\big(r^2+y_j^2\big) \big(r^2-y_i^2\big) + 2m s^{ 2} r\Big) dy_i \wedge \Big(\mathcal A_i - \frac{H-1}{H} \mathcal A\Big) \bigg]\,, \nn 
\end{align}
where no flat piece is needed: the regularity condition for the two-form, the vanishing of the pullback of $\mathrm{i}_\xi A_{(2)}$ to the horizon section $r=r_+$, is already satisfied by \eqref{eq:A2form6}, since at $r=r_+$ the $dy_i$ legs of $\mathrm{i}_\xi A_{(2)}$ vanish by $\mathrm{i}_\xi\big(\mathcal A_i-\frac{H-1}{H}\mathcal A\big)\big|_{r=r_+}=0$ and the remaining $dr$ leg pulls back to zero. 

The coordinate ranges and periodic identifications are
\begin{align}
	(\tau,\phi_i)\sim(\tau+\beta,\phi_i-i\Omega_i\beta)\,,\quad r\ge r_+\,,\quad -a_1 \leq y_1 \leq a_1 \leq y_2 \leq a_2 \,,\quad \phi_i\sim\phi_i+2\pi\,, \label{eq:ranges6}
\end{align}
for $0\le a_1\le a_2$ without loss of generality. The first periodic identification is the thermal cycle \eqref{eq:identification} with $n=2$, where $\beta$ and $\Omega_i$ are presented in \eqref{eq:thermo6}. The ranges of $y_{1,2}$ and $\phi_i$ are equivalent to \eqref{eq:mu} with $(n,\epsilon)=(2,1)$ through \eqref{eq:metric6def}.

\medskip

\noindent {\bf{Euclidean on-shell action.}} As in Subsection~\ref{sec:holo-renorm:AdS5}, the trace of the Einstein equation \eqref{eq:Einstein6} and the form-field equations \eqref{eq:formEOM6} reduce the bulk action \eqref{eq:action6} on shell to
\begin{align}
	S_{\rm bulk} \Big|_{\rm on\text-shell} &= - \frac{1}{16 \pi G_N} \int \bigg[\frac{1}{2} \mathcal V \star 1 + \frac{1}{2} g^2 (X_1 X_2)^{-1} \star A_{(2)} \wedge A_{(2)} -\frac{i}{6} g^2 A_{(2)} \wedge A_{(2)} \wedge A_{(2)} \nn\\
	&\quad + d \Big(-\frac{1}{4} \sum_{I=1}^2 X_I^{-2} \star  F_{(2)}^I \wedge A_{(1)}^I  + \frac{i}{4} \big(F_{(2)}^2 \wedge A_{(2)} \wedge A_{(1)}^1 + F_{(2)}^1 \wedge A_{(2)} \wedge A_{(1)}^2\big) \nn\\
	&\kern3em~ + \frac{1}{2} (X_1 X_2)^{2} \star  F_{(3)} \wedge A_{(2)} \Big)\bigg]\,, \label{eq:bulkonshell6}
\end{align}
where the flat connection again enters only through the total-derivative term, while the genuine bulk term now also contains the mass and cubic terms of the two-form and the Chern--Simons terms cancel only partially. The GHY term and the counterterms are \eqref{eq:GHY} and \eqref{eq:CT} with $d=6$. Although the scalar $X$ is non-trivial on the solution, the scalar dependence of the leading counterterm $S_{\text{CT-0}}$ through the superpotential drops out at the cutoff: $\mathcal W$ in \eqref{eq:potential6} is stationary at $X_I=1$, so that $\mathcal W-4g=\mathcal O\big((X-1)^2\big)=\mathcal O(\tir^{-6})$ by \eqref{eq:X6}, which vanishes against $\sqrt h=\mathcal O(\tir_\infty^5)$ --- in contrast to five dimensions, where $X_I-1=\mathcal O(\tir^{-2})$ leaves a finite remainder. For even $d$ there is no finite counterterm either, see the discussion below \eqref{eq:CT}. The renormalized on-shell action is \eqref{eq:Ireg} with $d=6$.

\medskip

\noindent {\bf{Asymptotically global AdS frame.}} For the AdS$_6$ black hole with $(n,\epsilon)=(2,1)$, the coordinate transformation \eqref{eq:GLPP} brings the associated Euclidean saddle \eqref{eq:metric6} to
\begin{align}
	ds^2 &= H^{\frac{1}{2}} \bigg[\big(1+g^2 \tir^2 \big) d\tau^2 + \frac{d\tir^2}{1+g^2 \tir^2} + \tir^2 \Big(  \sum_{i=1}^{3} d \tmu_i^2 +  \sum_{i=1}^{2} \tmu_i^2 d\phi_i^2\Big) \nn\\
	&\kern3em + \frac{2m\Upsilon(r)}{(1+g^2r^2)\big(V(r)-2m\Upsilon(r)\big) \tir^2 \sum_{i=1}^3 \frac{\Xi_i\tmu_i^2}{(r^2+a_i^2)^2}} \bigg(\frac{d\tir}{\tir} + \tir^2 \sum_{i=1}^3 \frac{\Xi_i\, \tmu_i d \tmu_i}{r^2+a_i^2}\bigg)^2 \nn\\
	&\kern3em +   \sum_{i=1}^2 \frac{Y_i}{\big(r^2+y_i^2\big)\big(y_i^2-y_j^2\big)} \Big(\frac{H-1}{H} \mathcal A - 2 \mathcal A_i\Big) \frac{H-1}{H} \mathcal A \nn\\
	&\kern3em + \frac{H^2\big(1+g^2r^2\big) \big(r^2+a_1^2\big)\big(r^2+a_2^2\big)- r\big(V(r)-2m\Upsilon(r)\big)}{H^2 \big(r^2+y_1^2\big)\big(r^2+y_2^2\big)}\mathcal A^2\bigg]\,, \label{eq:metric6new}
\end{align}
which manifestly approaches global EAdS$_6$ \eqref{eq:EAdS} at large $\tir$, with corrections of the form \eqref{eq:expansion} as follows from \eqref{eq:coordinates} with $(n,\epsilon)=(2,1)$. Explicitly, the new radial coordinate reads
\begin{align}
	\tir^2 = \frac{\big(1+g^2r^2\big) \big(1-g^2y_1^2\big) \big(1-g^2y_2^2\big)}{g^2 \Xi_1 \Xi_2} - \frac{1}{g^2} \label{eq:rtilde6}
\end{align}
by the coordinate transformation \eqref{eq:GLPP} and the $(y_1,y_2)$-parametrization of the original direction cosines $\mu_i$ of the Boyer--Lindquist-type coordinates as presented in \eqref{eq:metric6def}.

\medskip

\noindent {\bf{Renormalized on-shell action; bulk contributions.}} The bulk on-shell action \eqref{eq:bulkonshell6} is evaluated with the cutoff at $\tir=\tir_\infty$ following the steps of Subsection~\ref{sec:holo-renorm:AdS5}: the integrals over $(\tau,\phi_i)$ give $\beta(2\pi)^2$; the radial integral runs from the horizon $r=r_+$, located at
\begin{align}
	\tilde r(r_+, y_1, y_2)  = \Bigg[ \frac{\big(1+g^2r_+^2\big) \big(1-g^2y_1^2\big) \big(1-g^2y_2^2\big)}{g^2 \Xi_1 \Xi_2} - \frac{1}{g^2}\Bigg]^{\frac{1}{2}} \label{eq:rtildeplus6}
\end{align}
in the new radial coordinate, to the cutoff $\tir_\infty$, in practice over $r\in[r_+,r(\tir_\infty,y_1,y_2)]$ with $r(\tir,y_1,y_2)$ the inverse of \eqref{eq:rtilde6}; finally the $y_{1,2}$ integrals are performed. The result is 
\begin{align}
	S_{\rm bulk} \Big|_{\eqref{eq:metric6new}} &= \frac{g^2 \pi \beta}{3G_N} \tir_{\infty}^5 + \frac{5\pi m \beta s^{2}g^2\,\mathcal G^{(3)}_{00}}{2G_N} \tir_{\infty}^2 - \frac{5 \pi m  \beta s^{2}}{6 G_N \Xi_1 \Xi_2} + \frac{5 \pi m  \beta s^{2}\,\mathcal G^{(3)}_{00}}{2 G_N} - \frac{\pi \beta\, m^{2} s^{ 2} c^{ 2} r_+}{G_N \Xi_1 \Xi_2\, D_+} \nn\\
	&\quad - \frac{g^2 \pi \beta}{6G_N \Xi_1 \Xi_2\, D_+} \Big[2r_+ D_+^2 + m r_+^2 s^{2} D_+ + 2 m^{2} r_+ s^{4} \big(r_+^2 + a_1^2 + a_2^2\big) + 4m^{3} s^{6} \Big]\,, \nn\\
	\mathcal G^{(3)}_{00} &\equiv \frac{a_2 \tanh^{-1} (a_1 g)- a_1 \tanh^{-1} (a_2 g)}{a_1 a_2\big(a_1^2-a_2^2\big) g^3}\,, \label{eq:Sbulk6}
\end{align}
where $\mathcal G^{(3)}_{00}$ is one of the building-block integrals of Appendix~\ref{app:int6}. As in five dimensions, all terms of \eqref{eq:Sbulk6} come from the genuine bulk terms in the first line of \eqref{eq:bulkonshell6} except $-\frac{\pi\beta m^2s^2c^2r_+}{G_N\Xi_1\Xi_2D_+}=-\frac{\beta}{4}\Phi Q$, with $Q$ the total electric charge \eqref{eq:charges6}, which is the cutoff contribution of the total-derivative terms by Stokes' theorem in the regular gauge \eqref{eq:A6}.

\medskip

\noindent {\bf{Renormalized on-shell action; boundary contributions.}} The GHY term is evaluated with \eqref{eq:K}; the relevant coefficients of the expansion \eqref{eq:expansion} for the metric \eqref{eq:metric6new} are given as
\begin{align}
	\gamma_{\tir\tir}^{(0)} &=- \frac{3 m s^{ 2}}{g^2\mD^{\frac{3}{2}}}\,, \qquad \gamma_{\tir\tmu_a}^{(0)} = - \frac{4 m s^{ 2}}{g^2 \mD^{\frac{5}{2}}} \big(\Xi_a - \Xi_3\big)  \tmu_a\,, \nn\\
	\gamma_{\tir\tir}^{(1)} &= \frac{ m}{\mD^{\frac{7}{2}}}  \bigg[ \frac{2+7s^{2}}{g^4} + \frac{s^{2}}{g^2} \big(a_1^2+a_2^2\big) - s^{2} \Big(\sum_{i=1}^2a_i^2\tmu_i^2\Big)^2 \nn\\
	&\kern4em + \sum_{i=1}^2 \tmu_i^2 \Big( -\frac{4+27s^{2}}{2g^2}a_i^2 + \frac{13 s^{2}}{2} a_i^4 - s^{2} a_1^2 a_2^2\Big) \bigg]\,, \label{eq:gamma6}
\end{align}
with $\{x^i\}=\{\tau,\tmu_1,\tmu_2,\phi_1,\phi_2\}$. The trace of the perturbation $\alpha_{ij}$ of the induced metric \eqref{eq:hinduced} is
\begin{align}
	\alpha =  \bar{h}^{ij} \alpha_{ij} = \alpha^{(-3)} \tilde r_\infty^{-3} +  \alpha^{(-5)} \tilde r_\infty^{-5} +  \mathcal O(\tilde r_{\infty}^{-6})\,, \label{eq:alphatrace6}
\end{align}
where the expansion coefficients read
\begin{align}
	\alpha^{(-3)}  &= \frac{ 5 m s^{2}}{\mD^{\frac{3}{2}}}\,, \nn\\
	\alpha^{(-5)}  &= -  \frac{m}{\mD^{\frac{7}{2}}} \bigg[4g^2  s^{2} \Big(\sum_{i=1}^2 a_i^4  \tmu_i^2 - \Big(\sum_{i=1}^2 a_i^2  \tmu_i^2\Big)^2\Big) + \frac{2+ \big(4+3 (a_1^2 +a_2^2)g^2\big)s^{2}}{g^2} \nn\\
	&\kern5em + \sum_{i=1}^2 \frac{a_i^2}{2} \Big(-4+\big(-25 +11a_i^2 g^2-6a_j^2 g^2\big)s^{2}\Big) \tmu_i^2 \bigg]\,. \label{eq:alpha6}
\end{align}
Inserting the first-order terms of \eqref{eq:K} and \eqref{eq:sqrth}, which suffice for $d=6$ as discussed below \eqref{eq:Ricci}, together with the expressions \eqref{eq:gamma6} and \eqref{eq:alpha6} into the GHY term \eqref{eq:GHY} and performing the boundary integrals (Appendix~\ref{app:int6}) gives 
\begin{align}
	S_{\rm GHY} \Big|_{\eqref{eq:metric6new}}  &= -\frac{5g^2 \pi \beta}{3G_N} \tir_{\infty}^5 - \frac{4 \pi \beta}{3G_N} \tir_{\infty}^3 - \frac{25\pi m s^{2}\beta g^2\,\mathcal G^{(3)}_{00}}{2G_N} \tir_{\infty}^2 \nn\\
	&\quad - \frac{37\pi m s^{2}\beta\, \mathcal G^{(3)}_{00}}{2G_N} +  \frac{5\pi m \beta }{3G_N \Xi_1 \Xi_2} \Big[1+ \big(4-g^2\big(a_1^2+a_2^2\big)\big) s^{2} \Big]\,. \label{eq:SGHY6}
\end{align}

The counterterms \eqref{eq:CT} are evaluated likewise employing \eqref{eq:sqrth} and \eqref{eq:Ricci}, with $\bar{\mathcal R}=12/\tir_\infty^2$ and the quadratic combination in $S_{\rm CT\text{-}2}$ at leading order
\begin{align}
	\mathcal R_{ij} \mathcal R^{ij} - \frac{5}{16} \mathcal R^2 =  \bar{\mathcal R}_{ij} \bar{\mathcal R}^{ij}  - \frac{5}{16} \bar{\mathcal R}^2 +  \mathcal O(\tir_{\infty}^{-7})\,, \label{eq:quadratic6}
\end{align}
which gives
\begin{align}
	S_{\text{CT-0}} \Big|_{\eqref{eq:metric6new}}  &=  \frac{4g^2 \pi \beta}{3G_N} \tir_{\infty}^5 + \frac{2 \pi \beta}{3G_N} \tir_{\infty}^3 + \frac{10\pi m s^{2}\beta g^2\, \mathcal G^{(3)}_{00}}{G_N} \tir_{\infty}^2 -  \frac{ \pi \beta}{6 G_N g^2}  \tir_{\infty} \nn\\
	&\quad + \frac{13\pi m s^{2}\beta\, \mathcal G^{(3)}_{00}}{G_N} - \frac{4\pi m \beta }{3G_N \Xi_1 \Xi_2} \Big[1+ \big(4-g^2\big(a_1^2+a_2^2\big)\big) s^{2} \Big]\,, \nn\\
	S_{\text{CT-1}} \Big|_{\eqref{eq:metric6new}}  &=   \frac{2 \pi \beta}{3G_N} \tir_{\infty}^3 +   \frac{ \pi \beta}{3 G_N g^2}  \tir_{\infty} + \frac{3\pi m s^{2}\beta\, \mathcal G^{(3)}_{00}}{G_N}\,, \nn\\
	S_{\text{CT-2}} \Big|_{\eqref{eq:metric6new}}  &=   -   \frac{ \pi \beta}{6 G_N g^2}  \tir_{\infty}\,. \label{eq:SCT6}
\end{align}
Note that the scalar-dependent part of the leading counterterm $S_{\text{CT-0}}$ does not contribute, as explained below \eqref{eq:bulkonshell6}.

\medskip

\noindent {\bf{Renormalized on-shell action; total.}} Finally, substituting the bulk term \eqref{eq:Sbulk6}, GHY term \eqref{eq:SGHY6}, and the counterterms \eqref{eq:SCT6} into \eqref{eq:Ireg}, all divergences cancel, and we obtain the renormalized on-shell action for the six-dimensional Euclidean saddle \eqref{eq:metric6new} as
\begin{align}
	I_{\rm reg}^{(6)} &\equiv I_{\rm reg} \Big|_{\eqref{eq:metric6new}} = \beta \mathcal I^{(6)}\,, \label{eq:Ireg6}\\
	\mathcal I^{(6)} &= \frac{\pi m}{6 G_N \Xi_1 \Xi_2} \Big[2+3s^{2}-2g^2\big(a_1^2+a_2^2\big)s^{2}\Big] - \frac{\pi m^{2} s^{2} c^{2} r_+}{G_N \Xi_1 \Xi_2\, D_+} \label{eq:I6}\\
	&\quad - \frac{g^2 \pi}{6G_N \Xi_1 \Xi_2\, D_+} \Big[2r_+ D_+^2 + m r_+^2 s^{2} D_+ + 2 m^{2} r_+ s^{4} \big(r_+^2 + a_1^2 + a_2^2\big) + 4m^{3} s^{6} \Big]\,.\nn
\end{align}

\medskip

\noindent {\bf{Thermodynamics and QSR.}} The solution depends on the four parameters $(m,a_1,a_2,s)$. Its entropy, angular velocities, and temperature are \cite{Chow:2008ip,Cassani:2019mms} (with $G_N$ restored)
\begin{align}
	S &=  \frac{2\pi^2 D_+}{3 G_N \Xi_1 \Xi_2}\,, \qquad \Omega_i = \frac{a_i}{D_+}\Big[\big(1+g^2r_+^2\big) \big(r_+^2 + a_j^2\big) + 2 m g^2r_+ s^{2}\Big]\,, \nn\\
	T &= \frac{1}{\beta} = \frac{r_+\big(V'(r_+)-2m\Upsilon'(r_+)\big)}{4\pi D_+}\,, \label{eq:thermo6}
\end{align}
with the electrostatic potential $\Phi$ presented in \eqref{eq:Phi6}. The conserved charges are 
\begin{align}
	E &= \frac{2\pi m}{3G_N \Xi_1 \Xi_2} \bigg[\frac{1}{\Xi_1} + \frac{1}{\Xi_2} + s^{2} \Big(1+ \frac{\Xi_1}{\Xi_2} + \frac{\Xi_2}{\Xi_1}\Big)\bigg]\,, \nn\\
	Q &= \frac{2\pi m s c }{G_N \Xi_1 \Xi_2}\,, \qquad J_i = \frac{2\pi m a_i }{3 G_N \Xi_i^2 \Xi_j} \big(1+\Xi_j s^{2}\big)\,, \label{eq:charges6}
\end{align}
where $E$ is the energy, $J_i$ the angular momenta, and $Q=Q_1+Q_2$ the total electric charge of the two equal-charge $\mathrm{U}(1)$'s (so that $\Phi Q=\sum_I\Phi^IQ_I$). These quantities satisfy the first law \cite{Chow:2008ip}
\begin{align}
	dE = T\, dS + \sum_{i=1}^2 \Omega_i\, dJ_i + \Phi\, dQ\,, \label{eq:firstlaw6}
\end{align}
and the QSR reads \cite{Choi:2018fdc,Cassani:2019mms} 
\begin{align}
	I_{\rm QSR}^{(6)} = \beta E -S-\beta \sum_{i=1}^2 \Omega_i J_i -\beta \Phi Q\,. \label{eq:IQSR6}
\end{align}
Using the horizon condition $V(r_+)=2m\Upsilon(r_+)$ to eliminate $m$, we verify that the renormalized on-shell action \eqref{eq:Ireg6} satisfies the QSR exactly, with no Casimir offset and no scheme ambiguity, in accordance with the absence of finite counterterms in even bulk dimension:
\begin{align}
	I_{\rm reg}^{(6)} = I_{\rm QSR}^{(6)}\,. \label{eq:IregQSR6}
\end{align}
To the best of our knowledge, \eqref{eq:Ireg6}--\eqref{eq:I6} is the first evaluation of the holographically renormalized on-shell action of the general non-extremal rotating electrically charged AdS$_6$ black hole of \cite{Chow:2008ip}, which confirms the QSR \eqref{eq:IQSR6} independently. We close this subsection with a few remarks.
\begin{itemize}
	\item The relation \eqref{eq:IregQSR6} plays a crucial role in precision holography. In the supersymmetric limit, the on-shell action of the Euclidean saddle \eqref{eq:metric6}, obtained from its thermodynamic quantities through the QSR \eqref{eq:IQSR6} \cite{Choi:2018fdc,Cassani:2019mms}, has been compared successfully with the superconformal index of the dual five-dimensional SCFTs at large $N$ \cite{Choi:2019miv,Crichigno:2020ouj}, an agreement expected on universality grounds to hold for any holographic five-dimensional SCFT \cite{Bobev:2019zmz}.  The associated entropy function, which is related to the former by a Legendre transform, was proposed in \cite{Choi:2018fdc} (see also \cite{Hosseini:2019iad}) and has likewise been reproduced from the index \cite{Choi:2019miv,Crichigno:2020ouj}. On the gravitational side, however, the on-shell action was evaluated indirectly through the QSR \cite{Cassani:2019mms}. The first-principles evaluation \eqref{eq:Ireg6} and the resulting identification \eqref{eq:IregQSR6} therefore place the above holographic comparison on more solid ground, without resorting to black hole thermodynamics, in parallel with the seven-dimensional analysis of \cite{Bobev:2025xan}.
	
	\item The on-shell action of the supersymmetric Euclidean solution has recently been evaluated by equivariant localization in \cite{Couzens:2025ghx}, from the fixed-point data of the supersymmetric Killing vector alone, in agreement with the entropy function of \cite{Choi:2018fdc,Cassani:2019mms}; the computation of this subsection is complementary as it does not rely on supersymmetry.
	
	\item Holographic renormalization has been applied in \cite{Suh:2018qyv} to a different class of AdS$_6$ black holes, the static supersymmetric magnetically charged black holes of \cite{Suh:2018tul}, whose entropy is reproduced by the topologically twisted index of the dual field theories \cite{Hosseini:2018uzp,Crichigno:2018adf,Hosseini:2018usu}. For those solutions, the radial coordinate is already of the Fefferman--Graham type \eqref{eq:FG} up to a simple redefinition, so that the change of coordinates \eqref{eq:GLPP}, which organizes our derivation of \eqref{eq:IregQSR6}, is not needed there. 
	
	\item Finally, since the computation has been organized in the $\mathrm{U}(1)\times\mathrm{U}(1)$ theory, the on-shell form \eqref{eq:bulkonshell6}, the regularity conditions, and the general boundary analysis of Subsection~\ref{sec:holo-renorm:AdS5} apply without change to the single-rotation black holes with two independent charges of \cite{Chow:2011fh}; the explicit evaluation, with the solution data of \cite{Chow:2011fh} in place of \eqref{eq:metric6}--\eqref{eq:A2form6}, is left for future work. 
\end{itemize}

\section{Background Subtraction}\label{sec:bs}
In this section we compute the on-shell actions of the black holes of Section~\ref{sec:holo-renorm} and of the AdS$_7$ black hole of \cite{Bobev:2023bxl} by background subtraction: the divergences of the bulk and GHY terms are removed by subtracting the same quantities evaluated on global AdS, whose Euclidean period is fixed by matching the induced geometry on the cutoff surface \cite{Hawking:1995fd,Hawking:1998kw,Kunduri:2005zg,Chen:2005zj}; no counterterms are used.  We treat the AdS$_5$, AdS$_6$, and AdS$_7$ black holes in Subsections~\ref{sec:bs:AdS5}, \ref{sec:bs:AdS6}, and \ref{sec:bs:AdS7}, respectively. Intriguingly, at the constant-$\tir$ cutoff in the asymptotically global AdS frame of Section~\ref{sec:holo-renorm}, background subtraction reproduces neither the on-shell action obtained by holographic renormalization nor the one determined via the QSR, in any dimension. The mismatch has a universal structure in all dimensions, which we derive in Subsection~\ref{sec:bs:origin} from an exact relation between the two methods.

\subsection{AdS$_5$ black hole in $\mN=2$ STU gauged supergravity}\label{sec:bs:AdS5}

\noindent {\bf{Background subtraction at constant $\tir$.}} We cut both the black hole and the reference, global EAdS$_d$ \eqref{eq:EAdS}, at the same constant $\tir=\tir_\infty$ hypersurface in the asymptotically global AdS frame, and define the background-subtracted on-shell action (hence the subscript `BS') as
\begin{align}
	I_{\rm BS}^{(d)}  \equiv  \Big[S_{\rm bulk} + S_{\rm GHY}\Big]_{\mathcal M_d} - \Big[S_{\rm bulk} + S_{\rm GHY}\Big]_{\overline{\mathcal M}_d}\,,\qquad \overline{\mathcal M}_d\equiv\mathcal M_d\Big|^{\,r\ge r_+\ \to\ \tir\ge0}_{\,m=0,\ \beta\to\beta_0}\,, \label{eq:IBS}
\end{align}
universally for the Euclidean saddles $\mathcal M_5=\,$\eqref{eq:metric5new}, $\mathcal M_6=\,$\eqref{eq:metric6new}, and the seven-dimensional one $\mathcal M_7$ of \cite{Bobev:2023bxl}. Here $\overline{\mathcal M}_d$ denotes the reference, global EAdS$_d$ \eqref{eq:EAdS}: it is the $m=0$ member of the family $\mathcal M_d$ with the Euclidean time of period $\beta_0$ and the radial domain $0\le\tir\le\tir_\infty$. The period $\beta_0$ is fixed by matching the volumes of the two cutoff surfaces as 
\begin{align}
	\int_0^\beta d\tau \int_{S^{d-2}} \sqrt h \Big|_{\mathcal M_d,\ \tir = \tir_\infty}&= \int_0^{\beta_0} d\tau \int_{S^{d-2}} \sqrt h \Big|_{\overline{\mathcal M}_d,\ \tir = \tir_\infty} \nn\\
	&=\int_0^{\beta_0} d\tau \int_{S^{d-2}} \sqrt{\bar h} = \beta_0\, {\rm vol}\big(S^{d-2}\big)\, \tir_\infty^{d-2} \sqrt{1+g^2\tir_\infty^2}\,, \label{eq:volmatch}
\end{align}
where the induced metric of the reference at the cutoff is read off from \eqref{eq:EAdS} and \eqref{eq:hinduced} as
\begin{align}
	ds^2_{{\rm EAdS}_d}\Big|_{\tir=\tir_\infty} =\bh_{ij}dy^idy^j= \big(1+g^2 \tir_\infty^2 \big) d\tau^2  + \tir_\infty^2\, ds^2_{S^{d-2}}\,. \label{eq:refmetric}
\end{align}
For $d=5$, integrating $\sqrt h$ in \eqref{eq:sqrth5} with the measure \eqref{eq:measure5}, the matching condition \eqref{eq:volmatch} yields
\begin{align}
	\frac{\beta_0}{\beta} &= 1+ \frac{8ms_A \mathcal J^{(2)}_{00}}{3} \tilde r_\infty^{-2} + b_4 \ \tilde r_\infty^{-4} + \mathcal O(\tir_\infty^{-6})\,, \label{eq:beta05} \\
	b_4 &\equiv \int_0^1 d\tmu_1\, \tmu_1 \Big(\alpha^{(-4)} +\frac{\big(\alpha^{(-2)}\big)^2}{8}\Big)\,, \label{eq:b45}
\end{align}
where $\alpha^{(-2)}$, $\alpha^{(-4)}$ are given in \eqref{eq:alpha5}. 

From \eqref{eq:Sbulk5} and \eqref{eq:SGHY5}, we already have
\begin{align}
	\Big[S_{\rm bulk} + S_{\rm GHY}\Big]_{\mathcal{M}_5}  &= -\frac{3g^2 \pi \beta}{4G_N} \tilde r_{\infty}^4 - \frac{\pi \beta}{G_N} \bigg(\frac{3}{4} + 2 g^2 m s_A \mathcal J^{(2)}_{00}\bigg)\tilde r_{\infty}^2 + S_{{\text{bulk}}}^{{\text{fin}}}+S_{{\text{GHY}}}^{{\text{fin}}}\,, \label{eq:bulkGHY5}
\end{align}
while the reference contribution follows from the bulk and GHY on-shell actions on global EAdS$_5$ with $0\le\tir\le\tir_\infty$ and period $\beta_0$ as
\begin{align}
	S_{\rm bulk}\Big|_{\overline{\mathcal M}_5} &= \frac{\pi \beta_0 g^2}{4G_N} \tir_{\infty}^4\,, \qquad S_{\rm GHY}\Big|_{\overline{\mathcal M}_5} = -\frac{\pi \beta_0 g^2}{G_N} \tir_{\infty}^4 - \frac{3\pi \beta_0}{4G_N} \tir_{\infty}^2\,, \nn\\
	\Big[S_{\rm bulk} + S_{\rm GHY}\Big]_{\overline{\mathcal M}_5} &= -\frac{3 \pi \beta_0}{4G_N} \Big(g^2 \tir_{\infty}^4 + \tir_{\infty}^2\Big) \label{eq:reference5} \\
	&= -\frac{3g^2 \pi \beta}{4G_N} \tir_{\infty}^4 - \frac{\pi \beta}{G_N} \Big(\frac{3}{4} + 2 g^2 m s_A \mathcal J^{(2)}_{00}\Big)\tir_{\infty}^2 - \frac{3\pi\beta}{4G_N} \Big(\frac{8ms_A \mathcal J^{(2)}_{00}}{3}+g^2 b_4 \Big)\,,\nn
\end{align}
with \eqref{eq:beta05} used in the last line. Substituting \eqref{eq:bulkGHY5} and \eqref{eq:reference5} back into \eqref{eq:IBS} yields the background-subtracted on-shell action as
\begin{align}
	I_{{\text{BS}}}^{(5)} = S_{{\text{bulk}}}^{{\text{fin}}}+S_{{\text{GHY}}}^{{\text{fin}}} + \frac{3\pi\beta}{4G_N} \bigg(\frac{8ms_A \mathcal J^{(2)}_{00}}{3}+g^2 b_4 \bigg)\,. \label{eq:IBS5}
\end{align}

\medskip

\noindent {\bf{Comparison with holographic renormalization.}} Comparing the background subtraction result \eqref{eq:IBS5} with the one from holographic renormalization \eqref{eq:Ireg5} in the minimal scheme $\zeta=0$, we find
\begin{align}
	I_{\rm reg}^{(5)}\Big|_{\zeta=0} - I_{\rm BS}^{(5)} =\beta E_{{\rm AdS}_5} + \frac{\pi \beta g^2 m^2}{3G_N \Xi_1 \Xi_2} \big(s_A^2-3s_B\big) - \frac{\pi \beta m s_A \mathcal J^{(2)}_{00}}{2 G_N} \,. \label{eq:IregIBS5}
\end{align}
The three terms on the right-hand side of \eqref{eq:IregIBS5} are identified as follows. The first is the Casimir term $\beta E_{{\rm AdS}_5}$ in \eqref{eq:IregQSR5}, the renormalized on-shell action of the reference, which background subtraction removes by construction. Strictly speaking, the reference in the current formulation \eqref{eq:IBS} has period $\beta_0$, so that its renormalized on-shell action in the minimal scheme is $\beta_0E_{{\rm AdS}_5}$ up to terms vanishing as $\tir_\infty\to\infty$; since $\beta_0/\beta\to1$ in this limit by \eqref{eq:beta05}, it nevertheless coincides with $\beta E_{{\rm AdS}_5}$. The second is the finite contribution of the scalar-dependent part $\mathcal W-3g$ of the leading counterterm $S_{\text{CT-0}}$ in \eqref{eq:CT} with the superpotential \eqref{eq:potential5},
\begin{align}
	\Delta S_{{\rm CT}\text{-}0}&\equiv\lim_{\tir_\infty\to\infty}\frac{1}{8\pi G_N}\int d^4y\sqrt h\,\big(\mathcal W-3g\big)=\frac{\pi\beta g^2m^2}{3G_N\Xi_1\Xi_2}\big(s_A^2-3s_B\big)\,, \label{eq:DeltaCT05}
\end{align}
which vanishes if and only if the three charges are equal; it has no counterpart in background subtraction, where the scalars enter only through the bulk action. The last term is the part of the finite contribution of the curvature counterterm $S_{\text{CT-1}}$ in \eqref{eq:CT} that background subtraction does not reproduce. In Subsection~\ref{sec:bs:origin} all three terms are derived systematically, for generic $d$, from an exact relation between the two methods; there the last, geometric term is identified in every dimension with the $\delta\mathcal R$ term of $S_{\text{CT-1}}$, where $\delta\mathcal R$ is the first-order term defined in the expansion \eqref{eq:Ricci} of the boundary Ricci scalar, whereas the second, scalar term is specific to five dimensions. 

Trading $I^{(5)}_{\rm reg}|_{\zeta=0}$ in \eqref{eq:IregIBS5} for $I^{(5)}_{\rm QSR}$ by means of \eqref{eq:IregQSR5} and using the identification \eqref{eq:DeltaCT05}, we obtain an equivalent expression
\begin{align}
	I^{(5)}_{\rm BS}=\underbrace{I_{\rm reg}^{(5)}\Big|_{\zeta=0} - \beta E_{{\rm AdS}_5}}_{=I^{(5)}_{\rm QSR}} - \Delta S_{{\rm CT}\text{-}0}+\frac{\pi\beta ms_A\mathcal J^{(2)}_{00}}{2G_N}\,, \label{eq:IBSQSR5}
\end{align}
which clarifies that the background subtraction at the constant-$\tir$ cutoff reproduces neither the on-shell action from holographic renormalization nor the one determined via the QSR. 

\medskip

\noindent {\bf{Background subtraction at constant $\hat r$.}} The result \eqref{eq:IBSQSR5} should be contrasted with the classic background subtraction computation of \cite{Chen:2005zj} for the equal-charge solution of Chong, Cveti\v{c}, L\"u, and Pope (CCLP) \cite{Chong:2005hr}, which is recovered from \eqref{eq:metric5} by setting $s_1=s_2=s_3\equiv s$; see \cite{Kunduri:2005zg} for the analogous computation for the equal-rotation black hole of \cite{Cvetic:2004hs}. There the cutoff is placed at a constant value of the Boyer--Lindquist radius of \cite{Chong:2005hr,Chen:2005zj}, which is denoted by $r$ in those papers and by $\hat r$ here, to distinguish it from the radial coordinate $r$ of \eqref{eq:metric5}; the two are related by 
\begin{align}
	\hat r^2 = r^2 + 2ms^2\,. \label{eq:rhat5}
\end{align}
The reference is global EAdS$_5$ in the same spheroidal coordinates, \textit{i.e.}\ \eqref{eq:metric5} at $m=0$, cut at the same locus $\hat r=\hat r_\infty$ and integrated from the origin $\tir=0$, which the spheroidal chart represents as the surface $r^2=-\big(a_1^2\mu_1^2+a_2^2\mu_2^2-g^2a_1^2a_2^2\big)/\Delta_\theta\le0$ \cite{Chen:2005zj}. With $\beta_0$ fixed by the volume matching \eqref{eq:volmatch} on these surfaces, the subtraction \eqref{eq:IBS}, now performed at the constant-$\hat r$ surfaces, recovers the result of \cite{Chen:2005zj} for the general CCLP solution: the subtracted action coincides with the on-shell action \eqref{eq:IQSR5} determined via the QSR as
\begin{align}
	I_{\rm BS}^{(5)}\Big|_{\hat r,\ {\rm CCLP}}=I_{\rm QSR}^{(5)}\Big|_{\rm CCLP}=I_{\rm reg}^{(5)}\Big|_{\zeta=0,\ {\rm CCLP}}-\beta E_{{\rm AdS}_5}\,, \label{eq:CCLP5}
\end{align}
where the second equality is \eqref{eq:IregQSR5} at equal charges. In short, for the CCLP solution, holographic renormalization in the asymptotically global AdS frame and background subtraction agree up to the Casimir term, provided that the cutoff of the latter is placed at constant Boyer--Lindquist radius instead.  For the representative selected by the Boyer--Lindquist radial coordinate, holographic renormalization likewise differs from this background subtraction by a Casimir term, which then depends on the rotation parameters \cite{Bobev:2022ocx,Awad:2022exw}.

However, the choice of radial coordinate \eqref{eq:rhat5} for background subtraction of the CCLP solution, which brings the background-subtracted on-shell action into agreement with the one determined via the QSR as presented in \eqref{eq:CCLP5}, does not extend to a complete agreement for the general black hole with independent charges \eqref{eq:metric5}. To illustrate this point, we first note that the choice of radial coordinate \eqref{eq:rhat5} can be naturally extended to
\begin{align}
	\hat r^2\equiv r^2+\frac{2}{3}ms_A \label{eq:rhatgen5}
\end{align}
for the general black hole \eqref{eq:metric5}, which is motivated as the radial coordinate that absorbs the conformal factor $(H_1H_2H_3)^{1/3}$ of \eqref{eq:metric5} at large $\hat r$ and reduces to \eqref{eq:rhat5} at equal charges: up to the $K_I^2$ terms, the induced metric on a constant-$\hat r$ surface then takes the form of that of the reference, \eqref{eq:metric5} at $m=0$, in the coordinate $\hat r$ as
\begin{align}
	ds_5^2 &\supset (H_1 H_2 H_3)^{\frac{1}{3}} \bigg[- \frac{(1+g^2r^2)\Delta_\theta}{\Xi_1 \Xi_2} dt^2+\frac{\Sigma}{\Delta_\theta} d\theta^2+ \sum_{i=1}^2 \frac{r^2+a_i^2}{\Xi_i} \mu_i^2 d\phi_i^2\bigg] \\
	&= - \frac{(1+g^2\hat{r}^2)\Delta_\theta}{\Xi_1 \Xi_2} dt^2+\frac{\hat{r}^2+\sum_{i=1}^2a_i^2(1-\mu_i^2)}{\Delta_\theta} d\theta^2+ \sum_{i=1}^2 \frac{\hat{r}^2+a_i^2}{\Xi_i} \mu_i^2 d\phi_i^2+\mO(\hat{r}^{-2})\,. \nn
\end{align}
Implementing the background subtraction \eqref{eq:IBS} at constant-$\hat r$ surfaces for the black hole with independent charges \eqref{eq:metric5}, with $\beta_0$ fixed by the volume matching \eqref{eq:volmatch} on these cutoff surfaces, we find
\begin{align}
	I_{\rm BS}^{(5)}\Big|_{\hat r}=I_{\rm QSR}^{(5)}-\Delta S_{{\rm CT}\text{-}0}\,. \label{eq:IBSrhat5}
\end{align}
The $\mathcal J^{(2)}_{00}$ term in \eqref{eq:IBSQSR5} is now removed, but the scalar term \eqref{eq:DeltaCT05} remains, so that the complete agreement \eqref{eq:CCLP5} is special to equal charges, where the scalars are trivial. 

From a more fundamental perspective, the agreement between holographic renormalization and background subtraction observed in \eqref{eq:CCLP5} for the CCLP solution, modulo the Casimir term, is \textit{ad hoc} in that the radial cutoff of background subtraction has to be adjusted to the solution at hand: the asymptotically global AdS frame provides, for holographic renormalization, a representative of the boundary conformal class that is independent of the parameters in any dimension, whereas no systematic criterion for the radial cutoff of background subtraction is available, and the choice that works for the CCLP solution does not extend to independent charges. We return to this point in Subsection~\ref{sec:bs:origin} and in Section~\ref{sec:discussion}.

\subsection{AdS$_6$ black hole in $\mN=4$ $\SU(2)$ gauged supergravity}\label{sec:bs:AdS6}
\noindent {\bf{Background subtraction at constant $\tir$.}} We repeat the background subtraction computation of Subsection~\ref{sec:bs:AdS5} for the six-dimensional Euclidean saddle $\mathcal M_6=\,$\eqref{eq:metric6new} of Subsection~\ref{sec:holo-renorm:AdS6}, for which holographic renormalization reproduces exactly the on-shell action determined via the QSR as \eqref{eq:IregQSR6}. The background-subtracted action can be read off from the general formula \eqref{eq:IBS} with $d=6$ as
\begin{align}
	I_{\rm BS}^{(6)}  \equiv  \Big[S_{\rm bulk} + S_{\rm GHY}\Big]_{\mathcal M_6} - \Big[S_{\rm bulk} + S_{\rm GHY}\Big]_{\overline{\mathcal M}_6}\,. \label{eq:IBS:6d}
\end{align}
The first part of \eqref{eq:IBS:6d} is determined immediately by the sum of the bulk contribution \eqref{eq:Sbulk6} and the GHY contribution \eqref{eq:SGHY6}. On the reference $\overline{\mathcal M}_6$, global EAdS$_6$ \eqref{eq:EAdS} with $0\le\tir\le\tir_\infty$ and period $\beta_0$, the bulk and GHY on-shell actions are
\begin{align}
	S_{\rm bulk}\Big|_{\overline{\mathcal M}_6} = \frac{g^2 \pi \beta_0}{3G_N} \tir_{\infty}^5\,, \qquad S_{\rm GHY}\Big|_{\overline{\mathcal M}_6} = -\frac{5g^2 \pi \beta_0}{3G_N} \tir_{\infty}^5 - \frac{4 \pi \beta_0}{3G_N} \tir_{\infty}^3\,, \label{eq:reference6}
\end{align}
respectively, and the period $\beta_0$ is fixed by the volume matching \eqref{eq:volmatch}. Concretely, expanding $\sqrt h$ on the cutoff surface of the black hole as in \eqref{eq:sqrth}, with the trace $\alpha$ given in \eqref{eq:alphatrace6}--\eqref{eq:alpha6}, and performing the angular integrals with the building-block integrals of Appendix~\ref{app:int6}, we find 
\begin{align}
	\frac{\beta_0}{\beta} &= 1+ \frac{15 m  s^{2}\, \mathcal G^{(3)}_{00}}{2} \tir_\infty^{-3} \nn\\
	&\quad + \bigg[\frac{6 m  s^{2}\,\mathcal G^{(3)}_{00}}{g^2} - \frac{m}{g^2 \Xi_1 \Xi_2} \Big(1+\big(4-g^2\big(a_1^2+a_2^2\big)\big) s^{ 2} \Big)\bigg] \tir_\infty^{-5} + \mathcal O \big(\tir_\infty^{-6}\big)\,. \label{eq:beta06}
\end{align}
Substituting \eqref{eq:Sbulk6}, \eqref{eq:SGHY6}, \eqref{eq:reference6}, and \eqref{eq:beta06} into \eqref{eq:IBS:6d}, all divergences cancel and the background-subtracted on-shell action reads
\begin{align}
	I_{\rm BS}^{(6)} = \beta\,\mathcal I^{(6)} + \frac{2\pi m \beta s^{2}\,\mathcal G^{(3)}_{00}}{G_N}\,, \label{eq:IBS6}
\end{align}
with $\mathcal I^{(6)}$ the finite quantity defined in \eqref{eq:I6}.

\medskip

\noindent {\bf{Comparison with holographic renormalization.}} Comparing the background subtraction result \eqref{eq:IBS6} with the holographically renormalized on-shell action \eqref{eq:Ireg6}, the two differ only by the $\mathcal G^{(3)}_{00}$ term as
\begin{align}
	I_{\rm reg}^{(6)} - I_{\rm BS}^{(6)}  = -\frac{2\pi m\beta s^{2}\,\mathcal G^{(3)}_{00}}{G_N}\,, \label{eq:IregIBS6}
\end{align}
so that, replacing $I^{(6)}_{\rm reg}$ by $I^{(6)}_{\rm QSR}$ according to \eqref{eq:IregQSR6},
\begin{align}
	I^{(6)}_{\rm BS}=I^{(6)}_{\rm QSR}+\frac{2\pi m\beta s^{2}\,\mathcal G^{(3)}_{00}}{G_N}\,. \label{eq:IBSQSR6}
\end{align}
Background subtraction at constant $\tir$ thus fails to reproduce the on-shell action \eqref{eq:IQSR6} determined via the QSR in six dimensions as well. The mismatch is the single finite term in \eqref{eq:IBSQSR6}, the counterpart of the last term in \eqref{eq:IregIBS5}. In Subsection~\ref{sec:bs:origin} the mismatch between the two methods is derived for generic $d$ from an exact relation, which identifies this term with the $\delta\mathcal R$ term, defined in \eqref{eq:Ricci}, of the curvature counterterm $S_{\text{CT-1}}$ in \eqref{eq:CT}.

\subsection{AdS$_7$ black hole in $\mN=4$ maximal gauged supergravity}\label{sec:bs:AdS7}

Finally we consider the AdS$_7$ black hole of \cite{Bobev:2023bxl}, a solution of the $\mathrm{U}(1)\times\mathrm{U}(1)$ truncation \cite{Liu:1999ai} of the maximal $\mathrm{SO}(5)$ gauged supergravity in seven dimensions, which is a consistent truncation of eleven-dimensional supergravity on $S^4$ \cite{Nastase:1999cb,Nastase:1999kf}. The solution carries three independent rotation parameters and two independent electric charges. Its holographically renormalized on-shell action was computed in \cite{Bobev:2025xan} in the asymptotically global AdS frame defined by the coordinate transformation \eqref{eq:GLPP} with $(n,\epsilon)=(3,0)$, and satisfies the QSR in the scheme specified below in \eqref{eq:Ireg7}. Here we perform the background subtraction at constant $\tir$ for this Euclidean saddle, denoted by $\mathcal M_7$ in \eqref{eq:IBS}, using the bulk and GHY on-shell actions of \cite{Bobev:2025xan}. We use the shorthand
\begin{align}
	\Xi_i &\equiv 1-a_i^2g^2\,, \qquad \mathcal C_7 \equiv \sum_{i=1}^3 \frac{\log \Xi_i}{\prod_{j\neq i}\big(\Xi_i-\Xi_j\big)}\,, \nn\\
	\mathcal Z &\equiv 2 \sum_{i=1}^3 a_i^2 g^2 +8\prod_{i=1}^3 a_i g - \sum_{i=1}^3 \bigg(a_i^4g^4 - \frac{2}{a_i^2g^2} \Big(\prod_{j=1}^3 a_j g\Big)^2\bigg)\,, \nn\\
	\mathcal B &\equiv \Big(1+\frac{s_1^2+s_2^2}{2}\Big)\Big(1+\sum_{i=1}^3 \Xi_i\Big) - \frac{1}{4} (c_1-c_2)^2 \mathcal Z\,. \label{eq:shorthand7}
\end{align}

\medskip

\noindent {\bf{Background subtraction at constant $\tir$.}} The background-subtracted on-shell action is \eqref{eq:IBS} with $d=7$,
\begin{align}
	I_{\rm BS}^{(7)}  \equiv  \Big[S_{\rm bulk} + S_{\rm GHY}\Big]_{\mathcal M_7} - \Big[S_{\rm bulk} + S_{\rm GHY}\Big]_{\overline{\mathcal M}_7}\,, \label{eq:IBS:7d}
\end{align}
whose first part is the sum of the bulk and GHY on-shell actions of the black hole evaluated with the cutoff $\tir=\tir_\infty$ \cite{Bobev:2025xan},
\begin{align}
	S_{\rm bulk} \Big|_{\mathcal M_7} &= \frac{g^2 \pi^2 \beta}{8G_N} \tir_{\infty}^6 - \frac{3g^2 \pi^2 \beta m (s_1^2+s_2^2)\,\mathcal C_7}{10G_N} \tir_{\infty}^2 -\frac{3 \pi^2 \beta m (s_1^2+s_2^2)\,\mathcal C_7}{10G_N} \nn\\
	&\quad - \frac{\pi^2 \beta m\,\mathcal B}{8G_N \Xi_1 \Xi_2 \Xi_3} + \beta \mathcal I^{(7)}\,, \label{eq:Sbulk7}
\end{align}
and
\begin{align}
	S_{\rm GHY} \Big|_{\mathcal M_7} &= - \frac{3g^2 \pi^2 \beta}{4G_N} \tir_{\infty}^6 -\frac{5 \pi^2 \beta}{8G_N} \tir_{\infty}^4 + \frac{9g^2 \pi^2 \beta m (s_1^2+s_2^2)\,\mathcal C_7}{5G_N} \tir_{\infty}^2 \nn\\
	&\quad + \frac{14 \pi^2 \beta m (s_1^2+s_2^2)\,\mathcal C_7}{5G_N} +\frac{3\pi^2 \beta m\,\mathcal B}{4G_N \Xi_1 \Xi_2 \Xi_3}\,, \label{eq:SGHY7}
\end{align}
where the finite term $\beta\mathcal I^{(7)}$ in \eqref{eq:Sbulk7}, with $\mathcal I^{(7)}$ given in \cite[(3.18)]{Bobev:2025xan}, matches the on-shell action determined via the QSR from the thermodynamic quantities of \cite{Bobev:2023bxl} as 
\begin{align}
	I^{(7)}_{\rm QSR}=\beta E-S-\beta\sum_{i=1}^3\Omega_iJ_i-\beta\sum_{I=1}^2\Phi^IQ_I=\beta\mathcal I^{(7)}\,. \label{eq:IQSR7}
\end{align}
On the reference $\overline{\mathcal M}_7$, global EAdS$_7$ \eqref{eq:EAdS} with $0\le\tir\le\tir_\infty$ and period $\beta_0$, the bulk and GHY on-shell actions are
\begin{align}
	S_{\rm bulk}\Big|_{\overline{\mathcal M}_7} = \frac{g^2 \pi^2 \beta_0}{8G_N} \tir_{\infty}^6\,, \qquad S_{\rm GHY}\Big|_{\overline{\mathcal M}_7} = -\frac{3g^2 \pi^2 \beta_0}{4G_N} \tir_{\infty}^6 - \frac{5 \pi^2 \beta_0}{8G_N} \tir_{\infty}^4\,, \label{eq:reference7}
\end{align}
and the period $\beta_0$ is fixed by the volume matching \eqref{eq:volmatch}. Evaluating the left-hand side of \eqref{eq:volmatch} with the boundary volume element \cite[(3.23)]{Bobev:2025xan}, the asymptotic expansion \cite[(3.20)--(3.21)]{Bobev:2025xan}, and the boundary integrals of \cite[Appendix~B]{Bobev:2025xan}, one finds
\begin{align}
	\frac{\beta_0}{\beta} = 1- \frac{12 m (s_1^2+s_2^2)\,\mathcal C_7}{5} \tir_\infty^{-4} - \bigg[\frac{2 m (s_1^2+s_2^2)\,\mathcal C_7}{g^2} + \frac{m\,\mathcal B}{g^2 \Xi_1 \Xi_2 \Xi_3}\bigg] \tir_\infty^{-6} + \mathcal O \big(\tir_\infty^{-8}\big)\,. \label{eq:beta07}
\end{align}
Substituting \eqref{eq:Sbulk7}, \eqref{eq:SGHY7}, \eqref{eq:reference7}, and \eqref{eq:beta07} into \eqref{eq:IBS:7d}, all divergences cancel and we obtain
\begin{align}
	I_{\rm BS}^{(7)} = \beta \mathcal I^{(7)} - \frac{\pi^2 \beta m (s_1^2+s_2^2)\,\mathcal C_7}{4G_N} = I^{(7)}_{\rm QSR} - \frac{\pi^2 \beta m (s_1^2+s_2^2)\,\mathcal C_7}{4G_N}\,, \label{eq:IBS7}
\end{align}
with \eqref{eq:IQSR7} in the second equality.

\medskip

\noindent {\bf{Comparison with holographic renormalization.}} For comparison, we consider the holographically renormalized on-shell action \eqref{eq:Ireg} of the same seven-dimensional Euclidean saddle $\mathcal M_7$, evaluated in \cite{Bobev:2025xan} with the counterterms \eqref{eq:CT} and \eqref{eq:CTfin} for $d=7$, which reads
\begin{align}
	I_{\rm reg}^{(7)} \equiv I_{\rm reg} \Big|_{\mathcal M_7} = \beta \mathcal I^{(7)} - \frac{5 \pi^2 \beta}{128 g^4 G_N} \big(1-25600\, g^5 \zeta\big)\,. \label{eq:Ireg7}
\end{align}
Two remarks on the counterterms are in order.
\begin{itemize}
	\item The superpotential in the leading counterterm $S_{\text{CT-0}}$ can be replaced by its pure-gravity value, $\mathcal W\to5g$, without affecting the result since the scalars fall off fast enough in seven dimensions so that no finite scalar term analogous to \eqref{eq:DeltaCT05} arises; see the discussions below \eqref{eq:CT} and \eqref{eq:bulkonshell6}.
	
	\item The finite counterterm $S_{\text{CT-3}}$ in \eqref{eq:CTfin} with $n=\lfloor(d-1)/2\rfloor=3$ is equivalent to the one of \cite{Bobev:2025xan}, whose coefficient there corresponds to $g^5\zeta$ in our convention, while the logarithmic counterterm of \cite{Bobev:2025xan} vanishes on this saddle since the integrated anomaly vanishes.
\end{itemize}
The on-shell action \eqref{eq:Ireg7} obtained by holographic renormalization matches the background-subtracted one \eqref{eq:IBS7} neither in the minimal scheme ($\zeta=0$) nor in the scheme $\zeta=1/(25600\,g^5)$ of \cite{Bobev:2025xan}, in which it satisfies the QSR by \eqref{eq:IQSR7}:
\begin{align}
	I_{\rm reg}^{(7)}\Big|_{\zeta=1/(25600\,g^5)} -  I_{\rm BS}^{(7)} = I_{\rm reg}^{(7)}\Big|_{\zeta=0} -  I_{\rm BS}^{(7)}-\beta E_{{\rm AdS}_7}=\frac{\pi^2 \beta m (s_1^2+s_2^2)\,\mathcal C_7}{4G_N}\,, \label{eq:IregIBS7}
\end{align}
where $E_{{\rm AdS}_7}$ is the Casimir energy of global AdS$_7$ given by \cite{Gibbons:2005jd,Bobev:2025xan}
\begin{align}
	E_{{\rm AdS}_7}=-\frac{5\pi^2}{128g^4G_N}\,. \label{eq:Casimir7}
\end{align}
As in five dimensions, \eqref{eq:Casimir5}, $\beta E_{{\rm AdS}_7}$ is the on-shell action of global EAdS$_7$ obtained by holographic renormalization in the minimal scheme. The $\mathcal C_7$ term on the right-hand side of \eqref{eq:IregIBS7} is the seven-dimensional counterpart of the $\mathcal J^{(2)}_{00}$ and $\mathcal G^{(3)}_{00}$ terms in \eqref{eq:IregIBS5} and \eqref{eq:IregIBS6}. The universal structure shared by the three mismatches \eqref{eq:IregIBS5}, \eqref{eq:IregIBS6}, and \eqref{eq:IregIBS7} is investigated in the following Subsection~\ref{sec:bs:origin}.

\subsection{Universal structure of the mismatch}\label{sec:bs:origin}

The differences between the on-shell actions obtained by holographic renormalization in the minimal scheme $\zeta=0$, from Section~\ref{sec:holo-renorm} and \cite{Bobev:2025xan}, and the background-subtracted on-shell actions \eqref{eq:IBS} of Subsections~\ref{sec:bs:AdS5}--\ref{sec:bs:AdS7}, both evaluated in the asymptotically global AdS frame, are collected as
\begin{subequations}\label{eq:IregIBSall}
\begin{align}
	I_{\rm reg}\Big|_{\eqref{eq:metric5new},\ \zeta=0}-I_{\rm BS}\Big|_{\eqref{eq:metric5new}}&=\beta E_{{\rm AdS}_5}+\Delta S_{{\rm CT}\text{-}0}-\frac{\pi\beta ms_A\,\mathcal J^{(2)}_{00}}{2G_N}\,, \label{eq:IregIBSall5}\\
	I_{\rm reg}\Big|_{\eqref{eq:metric6new}}-I_{\rm BS}\Big|_{\eqref{eq:metric6new}}&=-\frac{2\pi\beta ms^2\,\mathcal G^{(3)}_{00}}{G_N}\,, \label{eq:IregIBSall6}\\
	I_{\rm reg}\Big|_{\mathcal M_7,\ \zeta=0}-I_{\rm BS}\Big|_{\mathcal M_7}&=\beta E_{{\rm AdS}_7}+\frac{\pi^2\beta m(s_1^2+s_2^2)\,\mathcal C_7}{4G_N}\,, \label{eq:IregIBSall7}
\end{align}
\end{subequations}
where the backgrounds are made explicit in the subscripts as in \eqref{eq:IBS}, $I_{\rm reg}|_{\mathcal M_d}=I^{(d)}_{\rm reg}$ and $I_{\rm BS}|_{\mathcal M_d}=I^{(d)}_{\rm BS}$.

\medskip

The difference \eqref{eq:IregIBSall} can be understood systematically by analyzing an exact relation between the two methods: since the bulk and GHY terms are common to both, \eqref{eq:Ireg} and \eqref{eq:IBS} give
\begin{align}
	I_{\rm reg}\Big|_{\mathcal M_d}-I^{(d)}_{\rm BS}=I_{\rm reg}\Big|_{\overline{\mathcal M}_d}+\sum_{\mathfrak a=0}^{n}\bigg(\Big[S_{{\rm CT}\text{-}\mathfrak a}\Big]_{\mathcal M_d}-\Big[S_{{\rm CT}\text{-}\mathfrak a}\Big]_{\overline{\mathcal M}_d}\bigg)\,, \label{eq:BSidentity}
\end{align}
where every term is evaluated at the same cutoff $\tir=\tir_\infty$ in the asymptotically global AdS frame before the limit $\tir_\infty\to\infty$ is taken. The right-hand side of \eqref{eq:BSidentity} accounts for the mismatch in \eqref{eq:IregIBSall} term by term, as follows.
\begin{itemize}
	\item The first term $I_{\rm reg}|_{\overline{\mathcal M}_d}$ is the renormalized on-shell action \eqref{eq:Ireg} of the reference, global EAdS$_d$ with period $\beta_0$; in the minimal scheme ($\zeta=0$) it is $\beta_0E_{{\rm AdS}_d}$, which reduces to the Casimir term $\beta E_{{\rm AdS}_d}$ of \eqref{eq:IregIBSall} because the volume matching condition \eqref{eq:volmatch} implies $\beta_0/\beta\to1$ as $\tir_\infty\to\infty$.
	
	\item For the counterterm differences in \eqref{eq:BSidentity}, recall that every counterterm in \eqref{eq:CT} and \eqref{eq:CTfin} is of the form $S_{{\rm CT}\text{-}\mathfrak a}=\frac{1}{8\pi G_N}\int d^{d-1}y\sqrt h\,\mathcal Q_{\mathfrak a}$ with a local density $\mathcal Q_{\mathfrak a}$ built from the induced metric and the scalars. On the reference background with the cutoff surface $S^1\times S^{d-2}$ \eqref{eq:refmetric} and trivial scalars, every density becomes a constant depending only on $\tir_\infty$, namely $\mathcal Q_{\mathfrak a}|_{\overline{\mathcal M}_d}\equiv\bar{\mathcal Q}_{\mathfrak a}$. Employing the volume matching condition \eqref{eq:volmatch}, we can then rewrite each difference as
	\begin{align}
		\Big[S_{{\rm CT}\text{-}\mathfrak a}\Big]_{\mathcal M_d}-\Big[S_{{\rm CT}\text{-}\mathfrak a}\Big]_{\overline{\mathcal M}_d}=\frac{1}{8\pi G_N}\int d^{d-1}y\sqrt h\,\big(\mathcal Q_{\mathfrak a}-\bar{\mathcal Q}_{\mathfrak a}\big)\Big|_{\mathcal M_d}\,. \label{eq:DeltaCT}
	\end{align}
	Given \eqref{eq:DeltaCT}, the counterterm differences can be analyzed as follows.
	\begin{itemize}
		\item For $\mathfrak a=0$, $\mathcal Q_0-\bar{\mathcal Q}_0=\mathcal W-(d-2)g=\mathcal O(\tir^{6-2d})$, since $\mathcal W$ is stationary at $X_I=1$ and $X_I-1=\mathcal O(\tir^{3-d})$. Since the boundary volume factor scales as $\sqrt{h}=\mathcal{O}(\tir_\infty^{d-1})$, the integral \eqref{eq:DeltaCT} with $\mathfrak a=0$ yields the finite term \eqref{eq:DeltaCT05} for $d=5$ and vanishes for $d=6,7$, as discussed below \eqref{eq:CT} and \eqref{eq:bulkonshell6}.
		
		\item For $\mathfrak a=1$, $\mathcal Q_1-\bar{\mathcal Q}_1=\big(\mathcal R-\bar{\mathcal R}\big)/\big(2(d-3)g\big)=\big(\delta\mathcal R+\mathcal O(\tir_\infty^{4-2d})\big)/\big(2(d-3)g\big)$ by \eqref{eq:Ricci}, so that \eqref{eq:DeltaCT} with $\mathfrak a=1$ reads $\frac{1}{16\pi G_N(d-3)g}\int d^{d-1}y\sqrt{\bar h}\,\delta\mathcal R+\mathcal O(\tir_\infty^{3-d})$ by \eqref{eq:sqrth}. Evaluating this term explicitly, we find
		\begin{subequations}\label{eq:DeltaCT1}
			\begin{align}
				\frac{1}{32\pi gG_N}\int d^4y\sqrt{\bar h}\,\delta\mathcal R\,\Big|_{\eqref{eq:metric5new}}&=-\frac{\pi\beta ms_A\,\mathcal J^{(2)}_{00}}{2G_N}\,, \label{eq:DeltaCT1a}\\
				\frac{1}{48\pi gG_N}\int d^5y\sqrt{\bar h}\,\delta\mathcal R\,\Big|_{\eqref{eq:metric6new}}&=-\frac{2\pi\beta ms^2\,\mathcal G^{(3)}_{00}}{G_N}\,, \label{eq:DeltaCT1b}\\
				\frac{1}{64\pi gG_N}\int d^6y\sqrt{\bar h}\,\delta\mathcal R\,\Big|_{\mathcal M_7}&=\frac{\pi^2\beta m(s_1^2+s_2^2)\,\mathcal C_7}{4G_N}\,, \label{eq:DeltaCT1c}
			\end{align}
		\end{subequations}
		in agreement with the remaining terms of \eqref{eq:IregIBSall}.
		
		\item For $\mathfrak a\geq 2$, including the finite counterterm \eqref{eq:CTfin} for odd $d$, the density is of degree $\mathfrak a$ in the boundary curvature, so that $\mathcal Q_{\mathfrak a}-\bar{\mathcal Q}_{\mathfrak a}=\mathcal O\big(\bar{\mathcal R}^{\mathfrak a-1}\delta\mathcal R\big)=\mathcal O(\tir_\infty^{3-d-2\mathfrak a})$. Combined with the volume factor $\sqrt{h}=\mO(\tir_\infty^{d-1})$, this counting ensures that \eqref{eq:DeltaCT} with $\mathfrak a\geq 2$ vanishes as $\tir_\infty^{2-2\mathfrak a}$. 
	\end{itemize}
\end{itemize}

\medskip

The relation \eqref{eq:BSidentity} with \eqref{eq:DeltaCT} identifies the mismatch between holographic renormalization and background subtraction, both implemented in the asymptotically global AdS frame, in terms of the ingredients of the former: the renormalized on-shell action of the reference and the counterterms. It does not provide a criterion for background subtraction itself, namely a necessary and sufficient condition under which background subtraction, with no counterterm added by hand, agrees with holographic renormalization up to the Casimir term and thereby with the on-shell action determined via the QSR, the quantity to be compared with the dual field theory. Note that incorporating counterterms --- in particular divergent ones such as the leading superpotential counterterm $S_{\text{CT-0}}$ and the curvature counterterm $S_{\text{CT-1}}$ in \eqref{eq:CT} --- is not in line with the philosophy of background subtraction, where divergences are to be removed exclusively by the reference contribution rather than by supplementing the action with boundary counterterms; if a suitable set of counterterms had to be invoked for the background-subtracted on-shell action to agree with the on-shell action determined via the QSR, there would be no point in adopting that formulation rather than proceeding with holographic renormalization from the outset. With this in mind, establishing background subtraction independently of the counterterms of holographic renormalization, for generic asymptotically AdS backgrounds including the rotating electrically charged AdS black holes considered here, is left for future work. One possible direction would be to make precise the original prescription of Hawking and Horowitz \cite{Hawking:1995fd}, which requires the solution and the reference to induce fields on the cutoff surface that agree ``to sufficient order'' for their difference not to contribute as the surface recedes to infinity; since holographic renormalization has been confirmed to work universally in the asymptotically global AdS frame for the black holes of interest, it would be very interesting to establish an equally universal prescription for background subtraction. We also refer the reader to \cite{Guo:2025ohn,Chen:2025ary,Xiao:2025icr,Li:2020spf,Ma:2024ynp} for recent discussions of related issues.

\section{Discussions} \label{sec:discussion}

In this paper we have evaluated the on-shell actions of the general non-extremal rotating electrically charged AdS$_5$ and AdS$_6$ black holes of gauged supergravity via holographic renormalization and confronted them, together with the AdS$_7$ black hole analysis of \cite{Bobev:2025xan}, with background subtraction. The decisive ingredient is the asymptotically global AdS frame, in which the cutoff surface approaches the Einstein static universe $S^1\times S^{d-2}$ for all values of the parameters: the renormalized on-shell actions satisfy the QSR exactly in six dimensions and, as in seven dimensions \cite{Bobev:2025xan}, up to the scheme-dependent Casimir term in five dimensions. The holographic comparisons that so far relied on the QSR are thereby placed on a first-principles footing. In contrast, background subtraction with the cutoff surface at constant $\tir$ in the same asymptotically global AdS frame reproduces neither the holographically renormalized on-shell action nor the on-shell action determined via the QSR. We have identified the mismatch exactly, in terms of the renormalized on-shell action of the reference and the finite parts of the counterterms of holographic renormalization, while leaving a self-contained formulation of background subtraction to future work. Below we present a list of open questions.

The first question is the extension to matter-coupled gauged supergravities in dimensions higher than five. In particular, the six-dimensional computation of Subsection~\ref{sec:holo-renorm:AdS6} was organized in the $\mathrm{U}(1)\times\mathrm{U}(1)$ gauged supergravity of \cite{Chow:2011fh}, which uplifts to massive type IIA supergravity \cite{Couzens:2022lvg}, and applies without change to the single-rotation black holes with two independent electric charges of that theory \cite{Chow:2011fh}, whose on-shell action remains to be evaluated explicitly. Moreover, the most general black hole of this theory with two independent rotations and two independent electric charges is not known; its construction along the lines of \cite{Chow:2008ip,Chow:2011fh} and the evaluation of its holographically renormalized on-shell action in the asymptotically global AdS frame would complete the six-dimensional counterpart of the five-dimensional analysis, including the comparison with the superconformal index of the dual five-dimensional SCFTs \cite{Choi:2019miv,Crichigno:2020ouj}.

A universal formulation of background subtraction is another interesting direction, as discussed at the end of Subsection~\ref{sec:bs:origin}: what is needed is a criterion for the cutoff surface, \textit{i.e.}\ for the radial coordinate at a constant value of which both the black hole and the reference are to be cut off. For the rotating electrically charged black holes considered here, the criterion has to make background subtraction with no counterterm added agree with holographic renormalization up to the Casimir term, and thereby with the on-shell action determined via the QSR. A first concrete task would be to make the ``sufficient order'' of agreement required by Hawking and Horowitz \cite{Hawking:1995fd} precise; the common relative order $\tir_\infty^{3-d}$ at which the induced metric and the scalars of the black holes considered here deviate from those of the reference, \textit{cf.}\ \eqref{eq:expansion} and the scalar fall-off noted below \eqref{eq:CT}, may provide a useful guide in this direction. See also \cite{Guo:2025ohn,Chen:2025ary,Xiao:2025icr,Li:2020spf,Ma:2024ynp} for recent related discussions.

Our analysis also bears on equivariant localization in supergravity \cite{BenettiGenolini:2023kxp,Martelli:2023oqk}, reviewed in \cite{Sparks:2026taw}, which reduces the bulk on-shell action of supersymmetric solutions to fixed-point data but does not by itself prescribe how the asymptotic boundary is treated. In practice the divergences are removed either by holographic renormalization as in the four-dimensional and Romans supergravity computations \cite{BenettiGenolini:2023kxp,BenettiGenolini:2024xeo,BenettiGenolini:2024lbj,Crisafio:2024fyc,Hristov:2021qsw,Hristov:2024cgj,Couzens:2025ghx} or by background subtraction as in the five- and seven-dimensional black hole computations \cite{Colombo:2025ihp,BenettiGenolini:2025icr,Colombo:2025yqy,Park:2025fon,Gaar:2026nqq,BenettiGenolini:2026hmz,Cassani:2026teb,BenettiGenolini:2026cdw}; both options are spelled out in \cite{BenettiGenolini:2025icr}.  The lesson of the present work is that, whichever of the two methods is adopted within equivariant localization, the treatment of the asymptotic boundary has to be specified unambiguously: the representative of the boundary conformal class for holographic renormalization and the cutoff surface for background subtraction. For the former a natural choice for black holes with conformal boundary $S^1\times S^{d-2}$ is the Einstein static universe \cite{Gibbons:2004ai,Gibbons:2005vp}, on which the dual index is formulated: the Casimir term and the finite counterterms are then independent of the parameters, as confirmed here for the general rotating electrically charged black holes.  For background subtraction the radial coordinate in which the cutoff is placed has to be specified for the formulation to be well defined, since the background-subtracted on-shell action depends on this choice as observed in Subsection~\ref{sec:bs:AdS5}. Clarifying these aspects would put the localization results in odd and even dimensions on an equally solid footing, free from such ambiguities.

Going beyond the two-derivative level is the next step for precision holography. Higher-derivative corrections to the on-shell actions of AdS$_4$ \cite{Bobev:2020egg,Bobev:2021oku,Hristov:2021qsw,Hristov:2024cgj} and AdS$_5$ \cite{Bobev:2021qxx,Bobev:2022bjm,Cassani:2022lrk,Cassani:2023vsa,Cassani:2024tvk,Cano:2024tcr,Ma:2024ynp,Gold:2025ttt,Hristov:2025ygn} black holes have been studied extensively in this context, recently also by equivariant localization \cite{Hristov:2024cgj,BenettiGenolini:2026qdm,Gaar:2026nqq}. The analysis of this paper should then be repeated with such higher-derivative terms included. On the holographic renormalization side the higher-derivative counterterms are known in many cases, \textit{e.g.}\ for the Gauss--Bonnet term, and have been employed in four and five dimensions \cite{Bobev:2020egg,Bobev:2021oku,Cassani:2022lrk,Cassani:2023vsa,Cassani:2024tvk}; what remains to be confirmed is that the asymptotically global AdS frame renders the finite counterterms and the Casimir term independent of the parameters with these terms included, as it does at two derivatives. On the background subtraction side, a prescription for the radial cutoff that resolves the ambiguity discussed here at two derivatives has to remain valid once higher-derivative terms are included; only then does background subtraction qualify as a well-established formulation. The higher-derivative analyses that already rely on background subtraction \cite{Bobev:2022bjm,Guo:2025muo,Chen:2025ary,Gaar:2026nqq} are the first cases such a prescription has to cover. The six- and seven-dimensional black holes are the natural next cases for both questions, once the leading higher-derivative corrections to the corresponding gauged supergravities are available; see \cite{Ozkan:2024euj} for the status of higher-derivative supergravities in diverse dimensions.

\acknowledgments
We are grateful to Nikolay Bobev, Marina David, Vasil Dimitrov, Rishi Mouland, and Jaeha Park for valuable discussions. We also thank Hassan ElSayed and Ioannis Papadimitriou for valuable correspondence on the first version of this paper. The work is supported by the National Research Foundation of Korea (NRF) grant funded by the Korean government (MSIT), Grant No.~RS-2024-00449284; by the Sogang University Research Grant No.~202410008.01; and by the Basic Science Research Program of the NRF funded by the Ministry of Education through the Center for Quantum Spacetime (CQUeST), Grant No.~RS-2020-NR049598.

\appendix

\section{Conventions}\label{app:conventions}
In a $d$-dimensional spacetime, the Hodge star operator acts on a $p$-form as
\begin{align}
	\star (dx^{\mu_1}\wedge\cdots\wedge dx^{\mu_p})=\frac{1}{(d-p)!}\epsilon_{\nu_{1}\cdots \nu_{d-p}}{}^{\mu_1\cdots \mu_p}dx^{\nu_{1}}\wedge\cdots\wedge dx^{\nu_{d-p}}\,, \label{eq:Hodge}
\end{align}
where the index placement, different from $\epsilon^{\mu_1\cdots \mu_p}{}_{\nu_{1}\cdots \nu_{d-p}}$ by the sign factor $(-1)^{p(d-p)}$, is the one for which $\star\omega_p\wedge\omega_p=\frac{1}{p!}\,\omega_{\mu_1\cdots\mu_p}\omega^{\mu_1\cdots\mu_p}\star1$ for every $p$, so that all kinetic terms in \eqref{eq:action5L} and \eqref{eq:action6} carry the correct sign. The totally anti-symmetric tensor is defined in the coordinate basis as 
\begin{equation}
	\epsilon_{\mu_1\cdots\mu_d}=\begin{cases}
		\sqrt{|g|} & (\text{even permutation}) \\
		-\sqrt{|g|} & (\text{odd permutation}) \\
		0 & (\text{otherwise})
	\end{cases}\,,\label{eq:epsilon}
\end{equation}
where $g$ denotes the determinant of the $d$-dimensional metric. The Hodge star operator satisfies the identity
\begin{align}
	\star (\star \omega_p)=(-1)^{p(d-p)+s}\omega_p\,,
\end{align}
where $s=0$ for Euclidean signature and $s=1$ for Lorentzian signature.

\medskip

The Wick rotation from Lorentzian to Euclidean signature, $t=-i\tau$, acts on the components of tensors as
\begin{align}
	X^0 \rightarrow -iX^0 \,,\qquad  X_0 \rightarrow iX_0\,, \label{eq:Wickcomp}
\end{align}
under which differential forms remain unchanged, while the Hodge star operator \eqref{eq:Hodge} transforms as
\begin{align}
	\star_L \omega_p \rightarrow -i \star \omega_p\,. \label{eq:HodgeWick}
\end{align}

\section{Building-block integrals}\label{app:building}
This appendix collects the angular integrals used in the main text.

\subsection{Integrals for the AdS$_5$ black hole} \label{app:int5}
The boundary integrals of Subsections~\ref{sec:holo-renorm:AdS5} and \ref{sec:bs:AdS5} are performed over the cutoff surface $S^1\times S^3$ of global EAdS$_5$ \eqref{eq:EAdS}, with the unit $S^3$ metric written in terms of the direction cosines $\tmu_{1,2}$ satisfying the constraint $\tmu_1^2+\tmu_2^2=1$ as
\begin{align}
	ds_{S^3}^2 = \frac{d\tmu_1^2}{1-\tmu_1^2} +  \tmu_1^2 d\phi_1^2 +  \tmu_2^2 d\phi_2^2\,. \label{eq:S3}
\end{align}
The boundary volume element reads $\sqrt{\bar h}=\tir^3\sqrt{1+g^2\tir^2}\,\tmu_1$ in this coordinate system, and therefore various boundary integrals appearing in Subsections~\ref{sec:holo-renorm:AdS5} and \ref{sec:bs:AdS5} take the form
\begin{align}
	\int d^4 y \sqrt{\bar h}\, \{\ldots\}= \big(2\pi\big)^2 \beta\, \tir^3 \sqrt{1+g^2 \tir^2} \int_{0}^1 d\tmu_1\, \tmu_1\, \{\ldots\}\,, \label{eq:measure5}
\end{align}
where the integrands are polynomials in $\tmu_i^2$ divided by integer powers of $\mD=\Xi_1\tmu_1^2+\Xi_2\tmu_2^2$ introduced in \eqref{eq:coordinates}, and hence linear combinations of the building-block integrals
\begin{align}
	\mathcal J_{j_1 j_2}^{(p)}= \int_0^1 d\tmu_1\, \tmu_1\, \frac{\tmu_1^{2j_1} \tmu_2^{2j_2}} {\mD^{\frac{p}{2}+j_1+j_2}}\qquad   \big(p \in 2\mathbb Z_{\geq0}\big)\,. \label{eq:J5}
\end{align}
Since $\mD=\Xi_2+\big(\Xi_1-\Xi_2\big)\tmu_1^2$ is linear in the integration variable $\tmu_1^2$, these are elementary integrals. The ones used in the main text are evaluated as
\begin{align}
	\mathcal J_{00}^{(0)} &= \frac{1}{2}\,, \qquad \mathcal J_{00}^{(2)} = \frac{1}{2g^2 \big(a_1^2-a_2^2\big)} \log\frac{\Xi_2}{\Xi_1}\,, \qquad \mathcal J_{00}^{(4)} = \frac{1}{2\Xi_1 \Xi_2}\,, \qquad \mathcal J_{00}^{(6)} = \frac{\Xi_1+\Xi_2}{4\Xi_1^2 \Xi_2^2}\,, \nn\\
	\mathcal J_{10}^{(4)} &= \frac{1}{4\Xi_1^2 \Xi_2}\,, \qquad \mathcal J_{01}^{(4)} = \frac{1}{4\Xi_1 \Xi_2^2}\,, \qquad \mathcal J_{11}^{(2)} = \frac{1}{2\big(\Xi_1-\Xi_2\big)^2} \bigg[\frac{\Xi_1+\Xi_2}{2\Xi_1 \Xi_2} - 2\mathcal J_{00}^{(2)}\bigg]\,. \label{eq:J5values}
\end{align}
%

\subsection{Integrals for the AdS$_6$ black hole} \label{app:int6}

The boundary integrals of Subsections~\ref{sec:holo-renorm:AdS6} and \ref{sec:bs:AdS6} live on the cutoff surface $S^1\times S^4$ of global EAdS$_6$ \eqref{eq:EAdS}. We parametrize the unit $S^4$ by two of the three direction cosines of \eqref{eq:mu} with $(n,\epsilon)=(2,1)$, eliminating the third one through $\tmu_3=\pm\sqrt{1-\tmu_1^2-\tmu_2^2}$, which turns the round metric into
\begin{align}
	ds_{S^4}^2 &= \frac{1-\tilde \mu_2^2}{1-\tilde \mu_1^2-\tilde \mu_2^2} d\tilde \mu_1^2 + \frac{2 \tilde \mu_1 \tilde \mu_2}{1-\tilde \mu_1^2-\tilde \mu_2^2} d\tilde \mu_1  d\tilde \mu_2 +  \frac{1-\tilde \mu_1^2}{1-\tilde \mu_1^2-\tilde \mu_2^2} d\tilde \mu_2^2 \nn\\
	&\quad +  \tilde \mu_1^2 d\phi_1^2 +  \tilde \mu_2^2 d\phi_2^2\,. \label{eq:S4}
\end{align}
The corresponding boundary volume element is $\sqrt{\bar h}=\tir^4\sqrt{1+g^2\tir^2}\,\tmu_1\tmu_2/\sqrt{1-\tmu_1^2-\tmu_2^2}$. It is worth emphasizing that a single chart $(\tmu_1,\tmu_2)$ covers only the hemisphere $\tmu_3\ge0$ out of the full range $-1\le\tmu_3\le1$; since every integrand encountered below depends on $\tmu_3$ through $\tmu_3^2$ alone, the two hemispheres contribute equally, and the boundary integrals of Subsections~\ref{sec:holo-renorm:AdS6} and \ref{sec:bs:AdS6} thereby reduce to
\begin{align}
	\int d^5 y \sqrt{\bar h}\, \{\ldots\}= 2\,\big(2\pi\big)^2 \beta\, \tir^4 \sqrt{1+g^2 \tir^2} \int_{\substack{0 \leq \tmu_1,\tmu_2 \\ \tmu_1^2 + \tmu_2^2 \leq 1}} d\tmu_1 d\tmu_2 \frac{ \tmu_1 \tmu_2}{\sqrt{ 1-\tmu_1^2- \tmu_2^2}}\, \{\ldots\}\,, \label{eq:measure6}
\end{align}
with the overall factor of $2$ accounting for the second hemisphere. The integrands are now polynomials in $\tmu_i^2$ divided by half-integer powers of $\mD=\Xi_1\tmu_1^2+\Xi_2\tmu_2^2+\tmu_3^2=1-g^2\big(a_1^2\tmu_1^2+a_2^2\tmu_2^2\big)$ from \eqref{eq:coordinates}, and hence linear combinations of
\begin{align}
	\mathcal G_{j_1 j_2}^{(p)}= \int_{\substack{0 \leq \tmu_1,\tmu_2 \\ \tmu_1^2 + \tmu_2^2 \leq 1}} d\tmu_1 d\tmu_2\, \frac{ \tmu_1 \tmu_2}{\sqrt{ 1-\tmu_1^2- \tmu_2^2}}\, \frac{\tmu_1^{2j_1} \tmu_2^{2j_2}} {\mD^{\frac{p}{2}+j_1+j_2}} \qquad \big(p\ \text{odd}\,,\ \text{or}\ p=j_1=j_2=0\big)\,, \label{eq:G6}
\end{align}
where the case $p=j_1=j_2=0$ is the volume integral. Since $\mD$ depends on the parameters only through $g^2a_i^2$, the identities $\partial_{a_i^2}\mD^{-s}=s\,g^2\tmu_i^2\,\mD^{-s-1}$ and $g\,\partial_g\mD^{-s}=2s\big(\mD^{-s-1}-\mD^{-s}\big)$ translate into the recurrence relations
\begin{align}
	\mathcal G_{00}^{(p+2)} = \mathcal G_{00}^{(p)} + \frac{g}{p} \frac{d}{dg} \Big(\mathcal G_{00}^{(p)}\Big) \,,\qquad  \mathcal G_{j_1 j_2}^{(p)} =\Bigg( \prod_{k=0}^{j_1+j_2-1} \frac{2}{\big(p+2k\big)g^2}\Bigg) \partial_{a_1^2}^{j_1} \partial_{a_2^2}^{j_2} \ \mathcal G_{00}^{(p)}\,, \label{eq:G6recursion}
\end{align}
which generate every $\mathcal G^{(p)}_{j_1j_2}$ with odd $p$ from $\mathcal G^{(1)}_{00}$. Only two elementary integrals are therefore needed,
\begin{align}
	\mathcal G_{00}^{(0)}= \frac{1}{3}\,, \qquad \mathcal G_{00}^{(1)}= \frac{a_1 \Xi_2 \tanh^{-1} (a_2 g)- a_2 \Xi_1 \tanh^{-1} (a_1 g)}{2 a_1 a_2\big(a_1^2-a_2^2\big) g^3}\,, \label{eq:G6values}
\end{align}
the former being ${\rm vol}(S^4)/8\pi^2$. The integrals used in the main text then follow from \eqref{eq:G6recursion}: the first relation gives
\begin{align}
	\mathcal G_{00}^{(3)} &= \frac{a_2 \tanh^{-1} (a_1 g)- a_1 \tanh^{-1} (a_2 g)}{a_1 a_2\big(a_1^2-a_2^2\big) g^3} \,, \nn\\
	\mathcal G_{00}^{(5)} &= \frac{1}{3 \Xi_1 \Xi_2}\,, \qquad \mathcal G_{00}^{(7)}= \frac{1}{15 \Xi_1 \Xi_2}\Big(1+\frac{2}{\Xi_1}+\frac{2}{\Xi_2}\Big)\,, \label{eq:G6values2}
\end{align}
and the second one yields
\begin{align}
	\mathcal G_{10}^{(5)} &= \frac{2}{15 \Xi_1^2 \Xi_2} \,,\qquad  \mathcal G_{01}^{(5)}= \frac{2}{15 \Xi_1 \Xi_2^2}\,, \nn\\
	\mathcal G_{20}^{(3)} &= \frac{1}{15\Xi_1^2  a_1^5 a_2  \big(a_1^2 -a_2^2\big)^3 g^7 } \Big[g a_1 a_2 \big(a_1^2 - a_2^2\big) \Big(9g^2 a_1^4 + 3a_2^2 -a_1^2 \big(7+5g^2a_2^2\big) \Big) \nn\\
	&\kern9em +  a_2  \Xi_1^2 \Big(15 a_1^4 - 10 a_1^2 a_2^2 +3a_2^4\Big) \tanh^{-1} (a_1g) - 8a_1^5 \Xi_1^2\tanh^{-1} (a_2g)\Big]\,, \nn\\
	\mathcal G_{02}^{(3)} &= \mathcal G_{20}^{(3)} \Big|_{a_1\leftrightarrow a_2}\,, \nn\\
	\mathcal G_{11}^{(3)} &= \frac{2}{15 g^7 \big(a_1^2-a_2^2\big)^3} \bigg[\frac{g  \big(a_1^2 - a_2^2\big) \big(a_1^2+a_2^2-a_1^4 g^2-a_2^4 g^2 \big)}{a_1^2 a_2^2 \Xi_1 \Xi_2} \nn\\
	&\kern9em - \frac{5a_1^2-a_2^2}{a_1^3} \tanh^{-1} (a_1 g) -\frac{a_1^2-5a_2^2}{a_2^3} \tanh^{-1} (a_2 g)\bigg]\,. \label{eq:G6values3}
\end{align}
%

\bibliography{AdS-BH}
\bibliographystyle{JHEP}


\end{document}